\documentclass[twocolumn]{aastex63}
\usepackage{natbib}
\usepackage{amsmath}
\usepackage{chngcntr}
\usepackage{hyperref}
\usepackage[figuresright]{rotating}
\usepackage{ctable}
\usepackage{float}
\usepackage{graphicx}
\usepackage{realboxes}
\usepackage{epsfig}
\usepackage{threeparttablex, tablefootnote}
\usepackage{subfigure}
\usepackage{graphicx}
\newcommand{\uat}[2]{\href{http://astrothesaurus.org/uat/#2}{#1 (#2)}}
\usepackage{lineno}
\usepackage{CJK}
\usepackage{booktabs}
\usepackage{multirow}
\usepackage{amsmath}

\newcommand{\cahk}{Ca \scriptsize{\uppercase\expandafter{\romannumeral2}} \normalsize H and K }

\newcommand{\ca}{Ca \scriptsize{\uppercase\expandafter{\romannumeral1}} \normalsize}

\shortauthors{Han et al.}

\begin{document}

\begin{CJK*}{UTF8}{gbsn}
\title{Impact of Spectral Resolution on $S$-index and Its Application to Spectroscopic Surveys}


\author{Henggeng Han}
\affil{Key Laboratory of Optical Astronomy, National Astronomical Observatories, Chinese Academy of Sciences, Beijing 100101, People's Republic of China}

\author{Song Wang$^\dagger$}
\affil{Key Laboratory of Optical Astronomy, National Astronomical Observatories, Chinese Academy of Sciences, Beijing 100101, People's Republic of China}
\affil{Institute for Frontiers in Astronomy and Astrophysics, Beijing Normal University, Beijing, 102206, People's Republic of China}
\email{$^\dagger$ Corresponding Author: songw@bao.ac.cn}

\author{Xue Li}
\affil{Key Laboratory of Optical Astronomy, National Astronomical Observatories, Chinese Academy of Sciences, Beijing 100101, People's Republic of China}
\affil{School of Astronomy and Space Science, University of Chinese Academy of Sciences, Beijing 100049, People's Republic of China}

\author{Chuanjie Zheng}
\affil{Key Laboratory of Optical Astronomy, National Astronomical Observatories, Chinese Academy of Sciences, Beijing 100101, People's Republic of China}
\affil{School of Astronomy and Space Science, University of Chinese Academy of Sciences, Beijing 100049, People's Republic of China}

\author{Jifeng Liu}
\affil{Key Laboratory of Optical Astronomy, National Astronomical Observatories, Chinese Academy of Sciences, Beijing 100101, People's Republic of China}
\affil{School of Astronomy and Space Science, University of Chinese Academy of Sciences, Beijing 100049, People's Republic of China}
\affil{Institute for Frontiers in Astronomy and Astrophysics, Beijing Normal University, Beijing, 102206, People's Republic of China}
\affil{New Cornerstone Science Laboratory, National Astronomical Observatories, Chinese Academy of Sciences, Beijing, 100012, People's Republic of China}

\begin{abstract}
Utilizing the PHOENIX synthetic spectra, we investigated the impact of spectral resolution on the calculation of $S$-indices. We found that for spectra with a resolution lower than $\approx$30,000, it is crucial to calibrate $S$-indices for accurate estimations. This is especially essential for low-resolution spectral observations.
We provided calibrations for several ongoing or upcoming spectroscopic surveys such as the LAMOST low-resolution survey, the SEGUE survey, the SDSS-V/BOSS survey, the DESI survey, the MSE survey, and the MUST survey.
Using common targets between the HARPS and MWO observations, we established conversions from spectral $S$-indices to the well-known $S_{\rm MWO}$ values, applicable to stars with [Fe/H] values greater than $-$1.
These calibrations offer a reliable approach to convert $S$-indices obtained from various spectroscopic surveys into $S_{\rm{MWO}}$ values and can be widely applied in studies on chromospheric activity.

\end{abstract}

\section{Introduction} 
\label{intro.sec}
The \cahk lines are two strong resonance lines in stellar optical spectra. The line wings could be modeled using local thermodynamic equilibrium (LTE), thereby unveiling the temperature structure of the photosphere. \citep{2004A&A...416..333R, 2012SoPh..280...83S, 2002A&A...389.1020R}. 
Meanwhile, the reversed structure and central absorption profile at the line center, caused by rising temperature and source function's decoupling from the Planck function, respectively, can serve as diagnosis of temperature structures of both lower and upper chromosphere \citep{1981ApJS...45..635V, 2013ApJ...772...90L, 2018A&A...611A..62B}.

Detailed modeling of the emission cores of \cahk lines suggests that they are sensitive to magnetic fields, which is also confirmed by solar observations \citep{1955ApJ...121..349B, 2013ApJ...764L..11D, 2024MNRAS.527.2940C}. Naturally they are excellent tracers of solar active regions \citep{2023ApJ...956L..10S, 2024MNRAS.527.2940C}, solar or stellar flares \citep{2009A&A...498..853R, 2024A&A...682A..46P} and the 11-year solar cycle \citep{1967ApJ...147.1106S, 2022AN....34323996D}.

Acting as pioneer, \cite{1968ApJ...153..221W} carried out a campaign to monitor the flux of the stellar \cahk lines using the 100-inch telescope at the Mount Wilson Observatory (MWO), which was equipped with a two-channel photometer, named as HKP-1. Later on \cite{1978PASP...90..267V} designed a four channel spectrometer named as HKP-2 and put it on the 60-inch telescope at MWO to avoid the instrumental effects of HKP-1. 

Initially, slits with full width at half maximum (FWHM) of 1.09 \AA \, were applied for observing dwarfs. Later on, the slits were replaced by new devices to allow the choice between 1 \AA \, and 2 \AA \, window \citep{1991ApJS...76..383D}, with the later is more suitable for observing giants considering the Wilson-Bappu effect, which leads to wider emission cores of \cahk lines compared to dwarfs \citep{1957ApJ...125..661W}. These observations provided an elite sample with long-term monitoring of stellar \cahk activity levels, confirming that the stellar \cahk lines could also vary quasi-periodically \citep{1978ApJ...226..379W}. 

\begin{figure}
\includegraphics[width=0.47\textwidth]{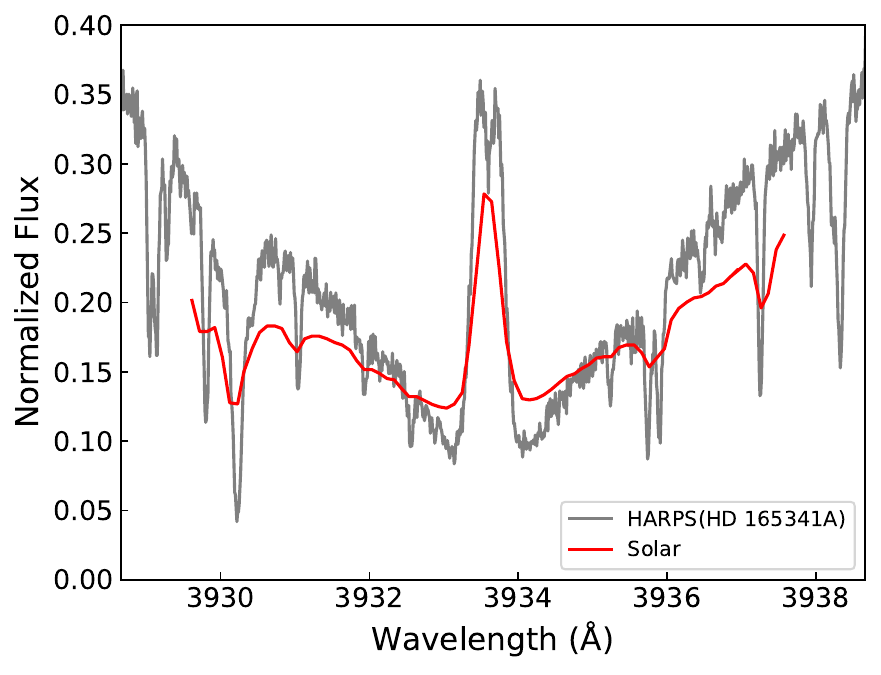}
\caption{Examples of observed emissions of Ca K lines of HD 165341A, a K0V star (gray) and spot region of the Sun (red). The solar spectrum was taken from \cite{2024MNRAS.527.2940C}, which was shifted for clarity.}
\label{observed}
\end{figure}

\begin{figure*}
\includegraphics[width=0.98\textwidth]{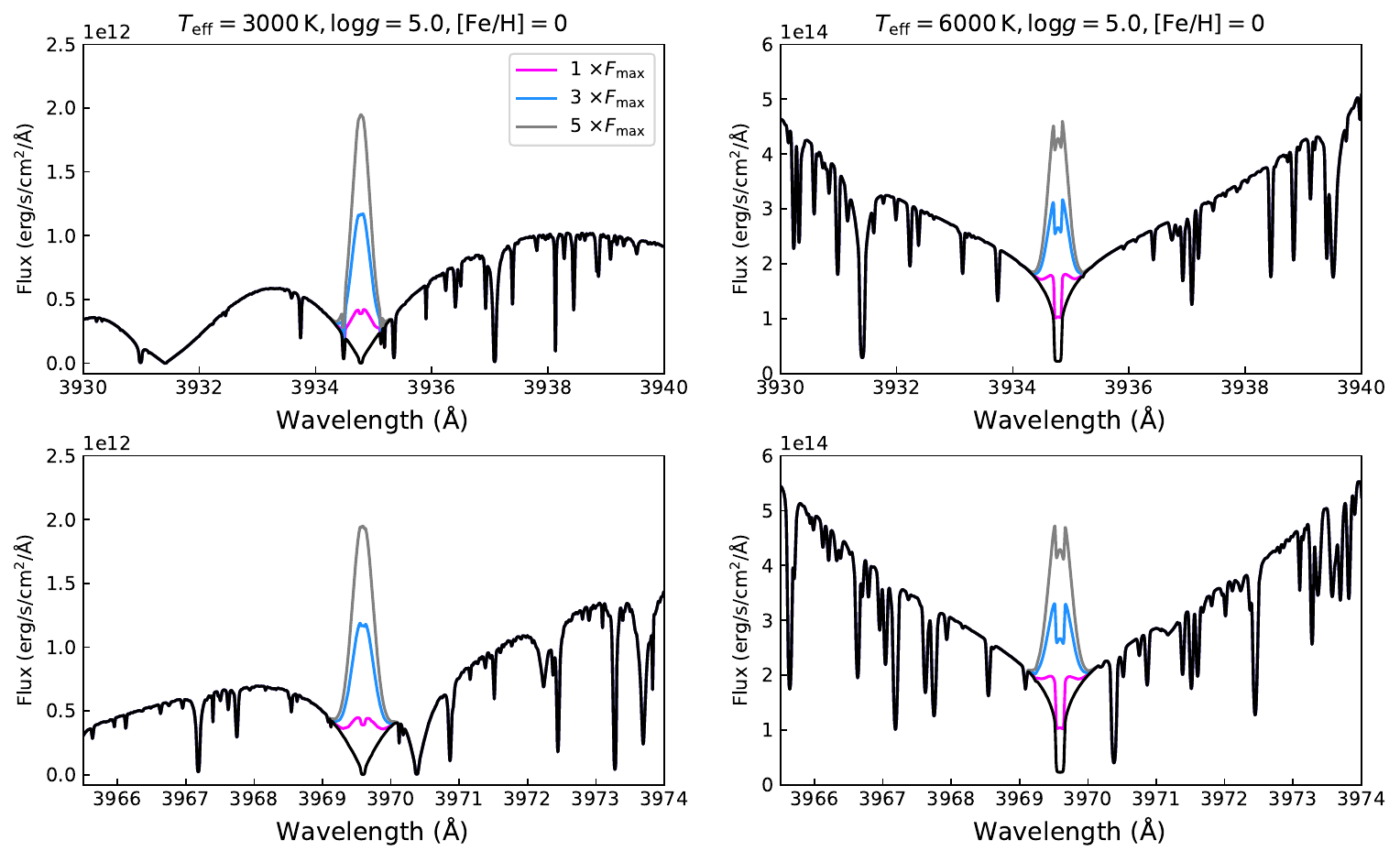}
\caption{Examples of the PHOENIX spectra and those with the Gaussian profiles added. Upper panels show the Ca K lines while lower panels show the Ca H lines. Different colors represent different emission levels.}
\label{spectra}
\end{figure*}

The chromospheric contribution to the \cahk lines was first defined as the observed flux of \cahk line cores with a basal flux to be subtracted, which was derived from a sample of inactive stars \citep{1968ApJ...153..221W}.
Later on, \cite{1978PASP...90..267V} proposed the well-known Mount Wilson $S$-index ($S_{\rm{MWO}}$) to quantify the chromospheric activity, defined as $S = \alpha(N_{\rm{H}} + N_{\rm{K}}) / (N_{\rm{R}} + N_{\rm{V}})$. $N_{\rm{H}}$ and $N_{\rm{K}}$ are the background-corrected counts in the H band centered at 3968.47 \AA \, and K band centered at 3933.664 \AA, respectively. $N_{\rm{R}}$ and $N_{\rm{V}}$ denote the background-corrected counts within the 20 \AA \, ranges of [3991.067 \AA, 4011.067 \AA] and [3891.067 \AA, 3911.069 \AA], respectively. 
$\alpha$ is the normalizing factor used to correct instrumental effects between HKP-1 and HKP-2 so that the $S$-indices are equal to the mean flux of the \cahk lines \citep{1978PASP...90..267V}. The $S_{\rm MWO}$ index has been widely used as an indicator of stellar chromospheric activity. 

Subsequent spectral observations calculated the spectra-based $S$-indices following a similar equation: $S = 8\alpha(H + K) / (R + V)$. \emph{H} and \emph{K} represent integrated flux corresponding to the H and K lines. The integration window was a triangle bandpass with a FWHM of 1.09 \AA. \emph{R} and \emph{V} are the integrated flux in the two 20 \AA \, rectangle reference bands at red and violet sides of H and K lines. The 8 is a correction factor for the longer exposure times of the V and R bandpass of the HKP-2 instrument.

However, the $\alpha$ value reported in the literature for different instruments vary significantly \citep[e.g.,][]{2006AJ....132..161G, 2007AJ....133..862H, 2018A&A...616A.108B}. 
Generally, the $\alpha$ value can be obtained by comparing the $S$-indices of the same stars observed by the MWO and other spectral surveys, allowing spectra-based $S$-indices to be converted to the Mount Wilson scale (i.e., $S_{\rm MWO}$) for consistent comparison.
Unfortunately, the stars observed by the MWO are typically bright, whereas current spectroscopic surveys increasingly target fainter stars. As a result, there are often few stars in common between the MWO observations and a given spectroscopic survey.
Therefore, it would be highly worthwhile to establish a straightforward and promising method to determine the conversion relations for different spectroscopic surveys.

Some instrumental effects including the spectral resolution, CCD response, ghosts and filter throughput could potentially influence the line profiles \citep[e.g.][]{2003PASP..115.1050S, 2020A&A...642A.182A, 2021A&A...648A.103D, 2021A&A...653A..43C, 2024A&A...682A..46P}, among which the spectral resolution plays an important role.
In this paper, we aim to use the PHOENIX synthetic spectra \citep{2013A&A...553A...6H} to investigate the impact of spectral resolution on the calculation of $S$-index. Furthermore, we will present the conversion relations between spectra-based $S$-indices and $S_{\rm MWO}$ across different spectral resolutions, especially for some wide-field spectroscopic surveys, including the LAMOST \citep{2012RAA....12.1197C, 2015RAA....15.1095L}, the SEGUE \citep{2009AJ....137.4377Y}, the SDSS-V \citep{2017arXiv171103234K}, the DESI \citep{2016arXiv161100036D}, and some upcoming surveys like the Maunakea Spectroscopic Explorer sky survey \citep[MSE;][]{2018arXiv181008695H} and the MUltiplexed Survey Telescope sky survey \citep[MUST;][]{2024arXiv241107970Z}.
The paper is organized as follows. In section 2 we introduce the data and method. Results and discussions are presented in Section 3.

\begin{figure}
\includegraphics[width=0.47\textwidth]{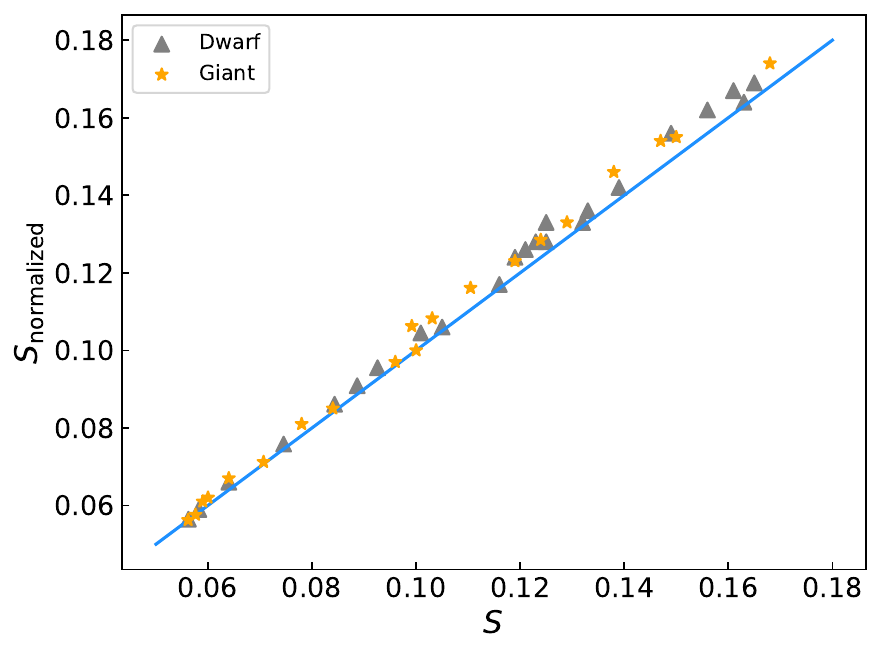}
\caption{Comparison between the $S$-indices calculated from unnormalized spectra and $S$-indices calculated from normalized spectra.}
\label{Comparison_nrom}
\end{figure}

\section{Data and Method}
\subsection{PHOENIX synthetic spectra}

PHOENIX high-resolution synthetic spectra have a resolution of 500,000. They are modeled under the local thermodynamic equilibrium (LTE), with non-LTE corrections applied for some specific lines including the \cahk lines \citep{2013A&A...553A...6H}. The library covers a range of effective temperatures ($T_{\rm{eff}}$) from 2300 K to 12,000 K, surface gravities (log$g$) from 0 to 6, and metallicities ([Fe/H]) from $-$4 to 1. In this work, we mainly focused on late-type dwarf and giant stars so that we only used the synthetic spectra with $T_{\rm{eff}}$ between 2500 K and 6500 K and log$g$ between 2 and 6. Meanwhile, all [Fe/H] grids have been included. No $\alpha$-enhancement was considered. We treated targets with log$g$ larger than 3.5 as dwarfs and those with log$g$ smaller than 3.5 as giants. Meanwhile, we marked targets with $6000 < T_{\rm{eff}} \leq 6500$ K as F-type stars, $5300 < T_{\rm{eff}} \leq 6000$ K as G-type stars, $4000 < T_{\rm{eff}} \leq 5300 $ K as K-type stars, and $T_{\rm{eff}} \leq 4000$ K as M-type stars.

\begin{figure*}
\includegraphics[width=1\textwidth]{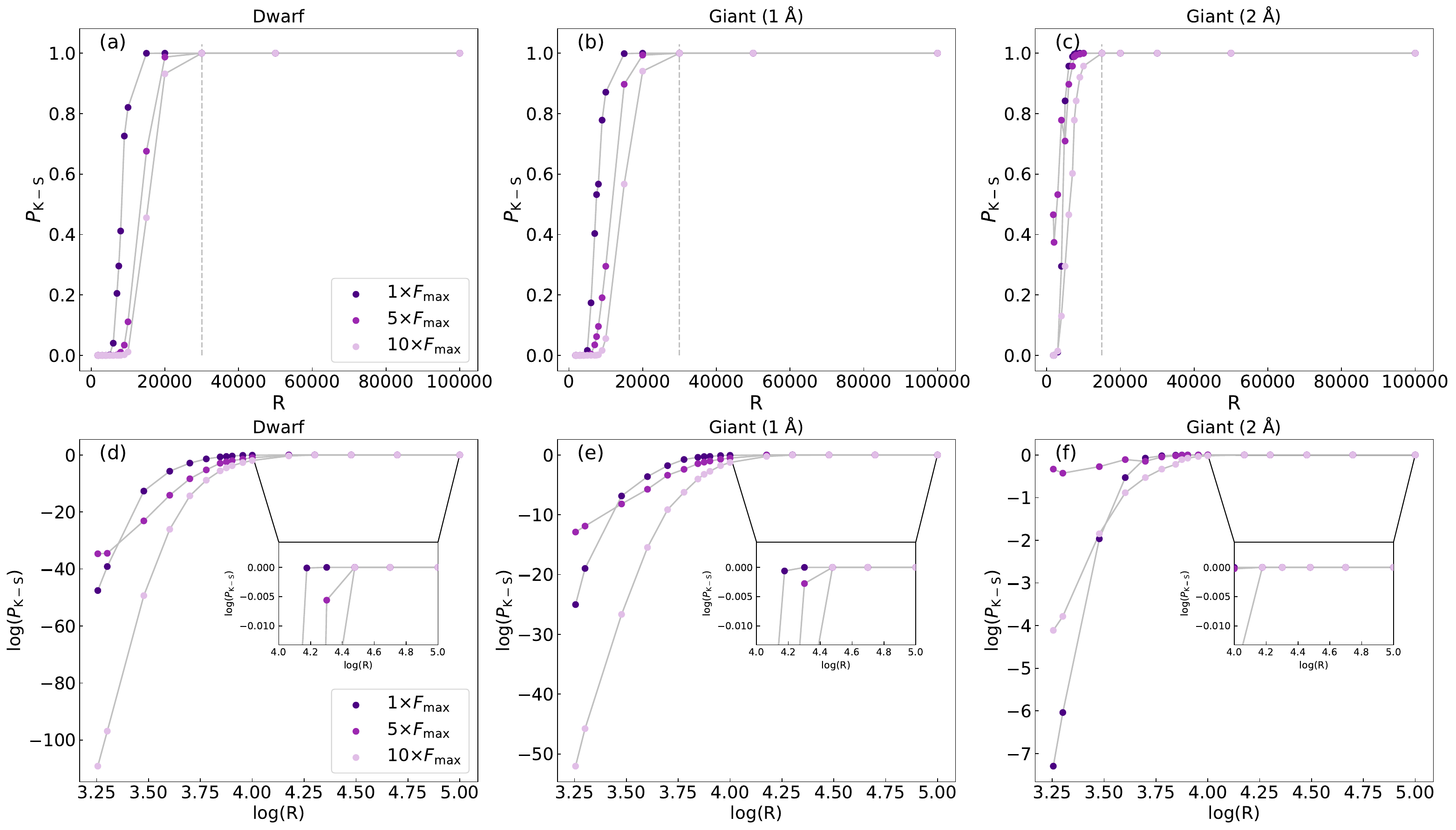}
\caption{Top panels: $P$-values from the K-S test ($P_{\rm{k-s}}$) for the $S_{\rm{O}}$ and $S_{\rm{R}}$ samples of dwarfs, the $S_{\rm{O}}$ and $S_{\rm{R}}$ samples of giants, and the $S_{\rm{O}}$ and $S_{\rm{R, 2A}}$ samples of giants, respectively. Different colors represent different strengths of the simulated emission cores. Gray vertical dashed lines mark the knee points where the two samples can be considered consistent. 
Bottom panels:  $P_{\rm{k-s}}$ values in logarithmic space.}
\label{ks}
\end{figure*}

To simulate spectra with chromospheric activity, we first constructed two normalized Gaussian profiles centered at 3934.78 \AA \, and 3969.59 \AA. In order to derive the full width at half maximum (FWHM) corresponding to stellar activity, we first conducted an estimation of the FWHM for the emission cores of the \cahk lines using HARPS spectra\footnote{http://archive.eso.org/wdb/wdb/adp/phase3\_main/form} for several active stars and the spectrum of solar spot region \citep{2024MNRAS.527.2940C}. In Figure \ref{observed} we plot two examples. The gray spectrum corresponds to HD 165341A (K0V), observed with HARPS while the red spectrum is the solar spectrum at the spot region \citep{2024MNRAS.527.2940C}. The results suggest that an FWHM of 0.35 \AA \, matches well with the observed spectra. As a result, in this work the FWHM of each Gaussian profile was set to be 0.35 \AA.  
These Gaussian profiles were then scaled by a group of factors, ranging from 1 to 10 with steps of 1, multiplying the maximum flux within the wavelength range between 3800 \AA \, and 4100 \AA, respectively, to represent different levels of chromospheric activity.
Finally, these Gaussian profiles were added to the PHOENIX spectra. Figure \ref{spectra} provides some examples.

\subsection{Calculation of $S$-index}

We calculated the $S$-index following 
\begin{equation}
    S = \frac{H + K}{R + V},
\end{equation}
excluding the factors of 8 and $\alpha$.
\emph{H} and \emph{K} represent the integrated fluxes corresponding to the \emph{H} and \emph{K} bands, centered at 3969.59 \AA \, and 3934.78 \AA, respectively. 
Note that these wavelengths mark the vacuum cores of the \cahk lines.
The integration window for these bands was a triangle bandpass with a FWHM of 1.09 \AA. \emph{R} and \emph{V} refer to the integrated fluxes within two 20 \AA \, rectangle reference bands on the red and violet sides of H and K lines, centered at 4001 \AA \, and 3901 \AA, respectively. 
In addition, considering the Wilson-Bappu effect, which indicates that the widths of the emission line cores of the \cahk lines are broader in giants compared to dwarfs \citep{1957ApJ...125..661W}, we also used a 2.18 \AA\ integration range for the $H$ and $K$ bands to calculate the $S$-indices for giants. It is worth noticing that a wider bandpass would wash out the weaker activity signal since more photospheric contribution from the wings would be brought in \citep{2022A&A...668A.174G, 2024A&A...682A..46P}.

\begin{figure*}
\includegraphics[width=0.98\textwidth]{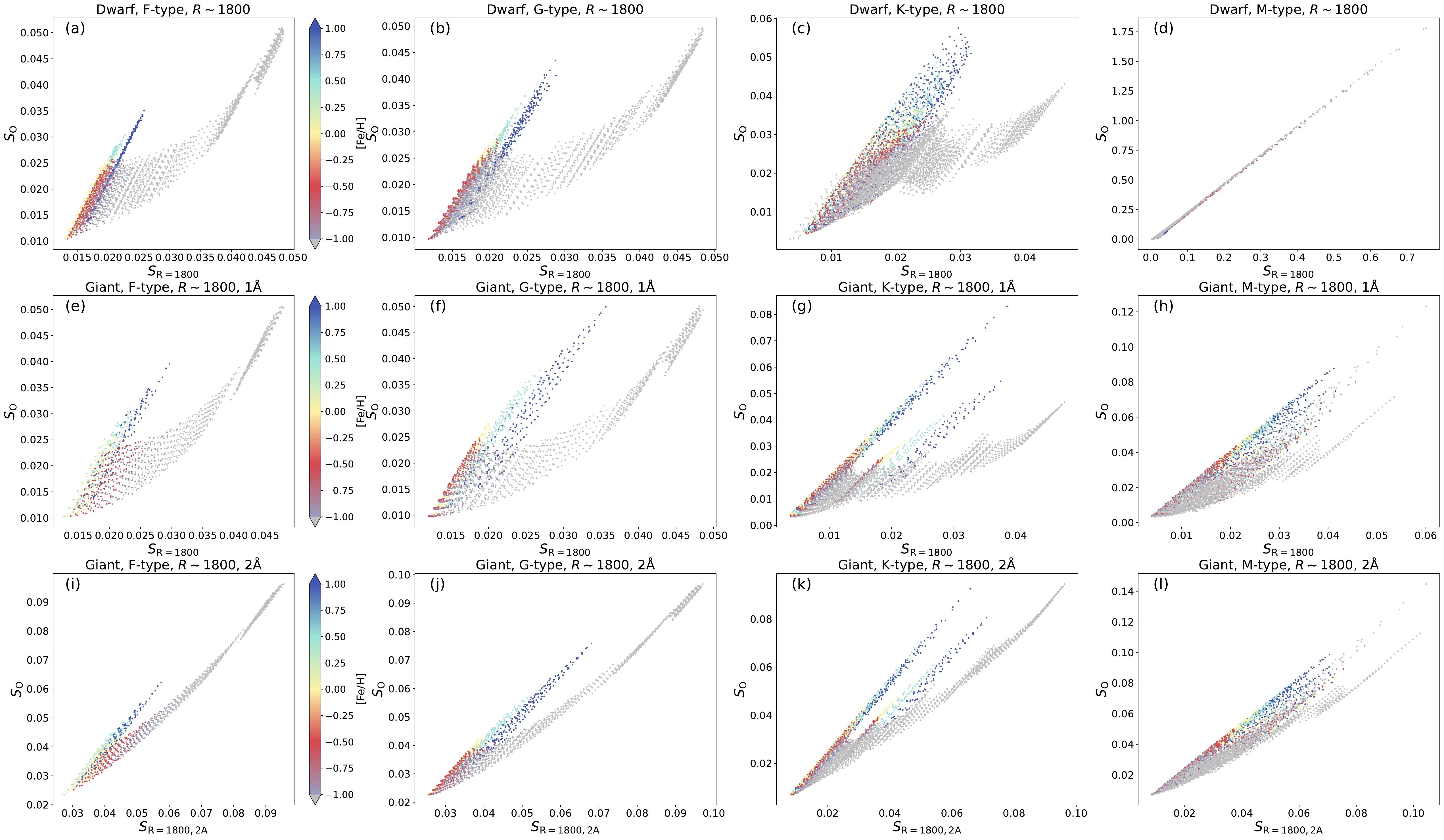}
\caption{Top panels: Relations between $S_{\rm{O}}$ and $S_{\rm{1800}}$ for different types of dwarfs. Different colors represent different [Fe/H]. Middle panels: Relations between $S_{\rm{O}}$ and $S_{\rm{1800}}$ for different types of giants.
Bottom panels: Relations between $S_{\rm{O}}$ and $S_{\rm{1800, 2A}}$ for different types of giants.}
\label{dwarf_giant_spec}
\end{figure*}

It should be noted that we used the original PHOENIX spectra (with flux units of erg/s/\AA/cm$^2$) for the calculations, while in many observations, normalized spectra were used \citep[e.g.][]{2004ApJS..152..261W, 2009A&A...493.1099S,  2024ApJ...977..138H}.
Therefore, we first conducted a check to see whether the $S$-indices calculated from normalized and unnormalized spectra are the consistent. 
We randomly selected a group of templates with various $T_{\rm{eff}}$ and log$g$ values and fixed [Fe/H] values of 0.0.
The comparison of the $S$-indices shows only a minor deviation, suggesting that both methods yield consistent results (Figure \ref{Comparison_nrom}).

We first calculated the $S$-indices for the original PHOENIX spectra with a resolution of $R=500,000$ and labeled them as $S_{\rm{O}}$. 
Then, the spectra were convolved to various resolutions of $R\sim1800$, 2000, 3000, 4000, 5000, 6000, 7000, 7500, 8000, 9000, 10,000, 15,000, 20,000, 30,000, 50,000, and 100,000 using a Gaussian window, and the corresponding $S$-indices were calculated, labeled as $S_{\rm{R}}$. 
For giants, the $S$-indices calculated using 2.18 \AA \, window were denoted as $S_{\rm{R, 2A}}$.
There are a total of 22,140 dwarf templates and 11,070 giant templates at each resolution.
These calculations will be used to investigate the influence of spectral resolution on the values of $S$-indices.

\section{Results and Discussions}

\subsection{The effects of spectral resolution}
\label{res.sec}

We used the standard nonparametric Kolmogorov-Smirnov (K-S) test to quantitatively assess at which resolution the $S_{\rm{O}}$ and $S_{\rm R}$ samples can be treated as the same.
Generally, the two samples can be regarded as different if the null hypothesis probability $P_{\rm K-S}$ is much smaller than one. 
To be conservative, here we treat they as the same if $P_{\rm K-S}$ $>$ 0.99.

It is obvious that at low resolutions, the $S_{\rm{O}}$ and $S_{\rm R}$ samples are significantly different (Figure \ref{ks}). Moreover, at the same resolution, the difference become more notable as the \cahk emission levels increase, indicating that stronger emission lines could amplify the impact of resolution.
Conversely, at high resolution, the two samples are quite consistent.

For both dwarfs and giants with $S$-indices calculated using a 1.09 \AA \, window, the knee point, where the $S_{\rm O}$ and $S_{\rm R}$ samples can be considered consistent, is around $R\sim30,000$. For giants with $S$-indices calculated with a 2.18 \AA \, window, the knee point is about $R\sim15,000$. 
We proposed that for spectra with $R\gtrsim30,000$, the effects of resolution on $S$-index estimation can be ignored (i.e., $S_{\rm O} = S_{\rm R}$). Our results are roughly consistent with \cite{2001A&A...374..733B} who proposed a 50,000 resolution threshold. The difference in threshold may be due to the different wavelength ranges. \cite{2001A&A...374..733B} used a wider wavelength range to investigate the influence of spectral resolution on line profile while we only focused on the narrow emission cores of \cahk lines.

\subsection{Conversion from $S_{\rm R}$ to $S_{\rm O}$}
\label{Sr2So.sec}

\begin{figure*}
\centering
\includegraphics[width=0.98\textwidth]{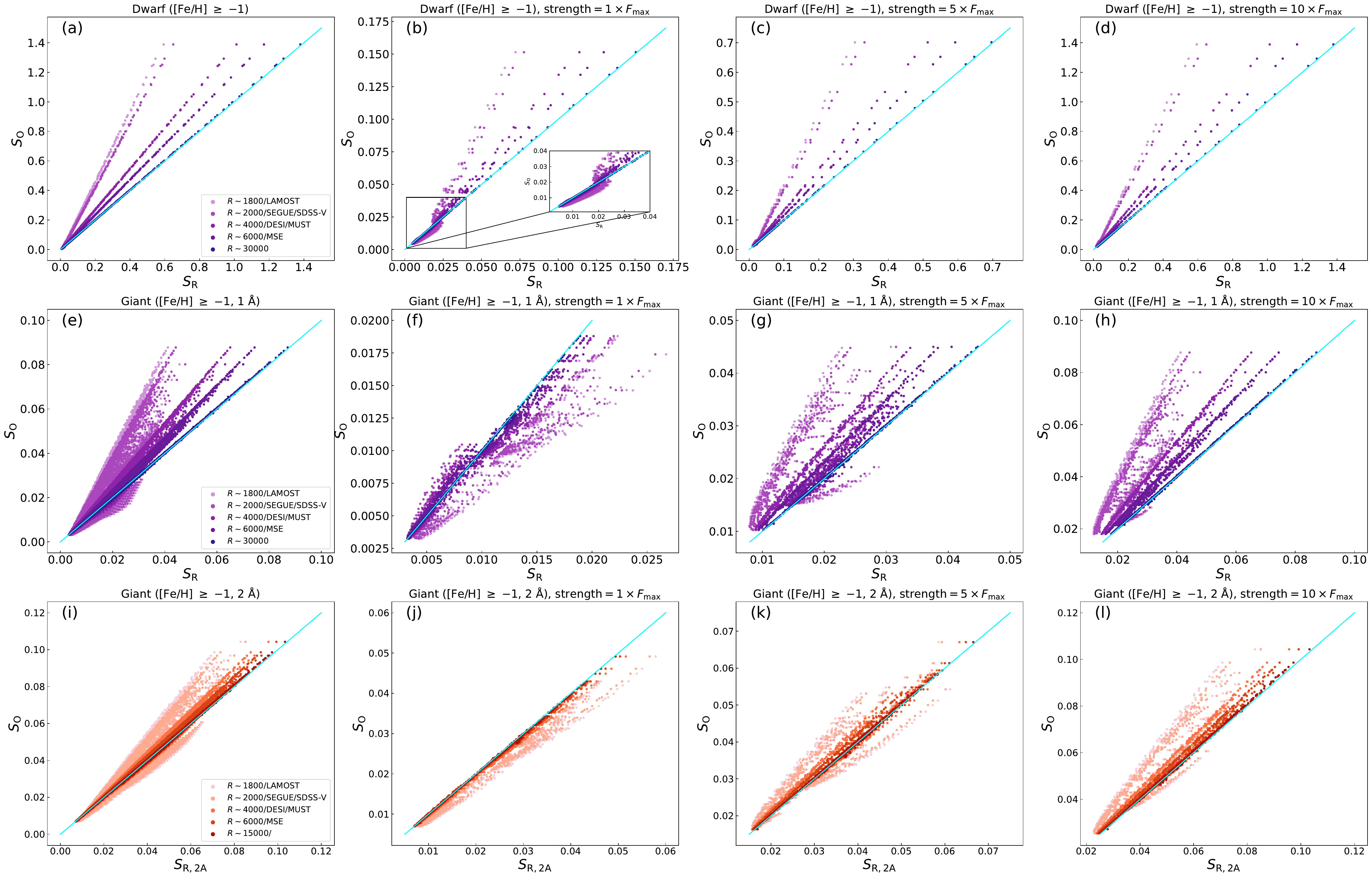}
\caption{Top panels: Relations between $S_{\rm{O}}$ and $S_{\rm{R}}$ for dwarfs with different line strengths. Different colors correspond to different spectral resolutions. Blue lines represent the $S_{\rm{O}} = S_{\rm{R}}$ line.
Middle panels: Relations between $S_{\rm{O}}$ and $S_{\rm{R}}$ for giants with different line strengths.
Bottom panels: Relations between $S_{\rm{O}}$ and $S_{\rm{R, 2A}}$ for giants with different line strengths. Blue lines represent the $S_{\rm{O}} = S_{\rm{R, 2A}}$ line.}
\label{res}
\end{figure*}

Considering the impact of spectral resolution on the calculation of $S$-indices, especially for low-resolution cases, one should convert the $S_{\rm R}$ values to $S_{\rm O}$ values to mimic the resolution effects.
Here we provide the calibrations for some large spectroscopic sky surveys, including the LAMOST low-resolution survey with $R\sim1800$ \citep{2012RAA....12.1197C}, the SEGUE survey with $R\sim2000$ \citep{2009AJ....137.4377Y}, the SDSS-V/BOSS survey with $R\sim2000$ \citep{2017arXiv171103234K}, the DESI survey with $R\sim4000$ \citep{2016arXiv161100036D}, and some upcoming sky surveys like the MSE survey with $R\sim6000 $ \citep{2018arXiv181008695H} and the MUST survey with $R\sim4000$ \citep{2024arXiv241107970Z}.

Figure \ref{dwarf_giant_spec} shows the comparison of $S_{\rm O}$ and $S_{\rm R=1800}$ for different types of stars.
For stellar templates with [Fe/H] $\geq -1$, the $S_{\rm O}$ and $S_{\rm R=1800}$ exhibit a linear relationship, whereas for templates with [Fe/H] $< -1$, the relationship becomes non-linear.
Since the integration window includes the line wings, the slope of the wings will affect the flux within the integrated region. For metal-rich stars, their broad wings have shallower slopes, making the integration less sensitive to slope variations. As a result, changes in metallicity, which affect the slopes, only have a minor impact on the integrated flux and thus lead to tight relations between $S_{\rm{R}}$ and $S_{\rm{O}}$. In contrast, metal-poor stars exhibit narrower line wings with steeper slopes. Consequently, the integrated flux becomes more sensitive to the wings, and changes in slopes caused by variations in [Fe/H] significantly influence the integrated flux. This results in large dispersions in the $S_{\rm{R}}$-$S_{\rm{O}}$ relations.

After excluding metal-poor stars, the relations are similar across different types of stars and the correlation is tighter for late-type stars. Therefore, we will focus our analysis on the relations for stellar templates with [Fe/H] $\geq -1$. In addition, our analysis reveals that excluding A- to M-type stars with [Fe/H] $< -1$ from the LAMOST DR10 catalog reduces the target sample by merely 4$\%$, suggesting that the excluding of these stars will not significantly influence the statistical analysis. For templates with [Fe/H] $\geq -1$, although there are small differences in the $S_{\rm{R}}$-$S_{\rm O}$ relations, we treat them as a single group.

Figure \ref{res} compares $S_{\rm O}$ and $S_{\rm R}$ for dwarfs with a 1.09 \AA\ widow, giants with a 1.09 \AA\ window and giants with a 2.18 \AA\ window, respectively.
Obviously, the deviation between $S_{\rm{O}}$ and $S_{\rm R}$ gradually increases as the spectral resolution decreases.
Additionally, for both dwarfs and giants, when the emission levels of \cahk lines are high, the relationships are very tight. However, when the emission levels are low, the relations would split into several branches in low-resolution cases (e.g., $R \leq 2000$).

For each resolution, a linear regression was applied to fit the relationship between $S_{\rm O}$ and $S_{\rm{R}}$ using templates with [Fe/H] $\geq -1$ with all the emission strengths:
\begin{equation}
    S_{\rm O} = k * S_{\rm R} + b.
\end{equation}
Furthermore, for low-resolution spectra, we divided the sample into two groups based on activity levels: a low-activity group ($S_{\rm R} <$ 0.02 for dwarfs and $S_{\rm R} <$ 0.01 for giants) and a high-activity group ($S_{\rm R} \geq$ 0.02 for dwarfs and $S_{\rm R} \geq$ 0.05 for giants). Individual linear regression was applied to each group.
The fitting results are listed in table \ref{tab:table1}. These individual fittings are recommended for $S$-index calibrations.
Additionally, for the low-activity group, we performed linear fittings for different types of stars, and the fitting results were found to be very similar. Finally, since the linear relations are tighter when calculating the $S$-indices using a 2.18 \AA \, compared to those that correspond to 1.09 \AA \, window (Figure \ref{res}), as reported by \cite{1991ApJS...76..383D, 2018MNRAS.480.2137S}, we suggested to use a 2.18 \AA \, window to calculate the $S$-indices for giants.

\subsection{Conversion from $S_{\rm R}$ to $S_{\rm MWO}$}

The $S$-indices represent the total emission of the \cahk lines, which includes the photospheric contribution unrelated to magnetic activity.
To address this, the $R_{\rm{HK}}^{'}$, which represents the ``flux excess" in the \cahk lines, was introduced as the standard chromospheric activity indicator.
It can be derived from the $S_{\rm MWO}$ through a series of complex steps \citep{1984ApJ...279..763N, 2013A&A...549A.117M}.
In the following analysis we will provide conversions from $S_{\rm R}$ to $S_{\rm MWO}$.

In Section \ref{Sr2So.sec}, we presented the conversions from $S_{\rm R}$ to $S_{\rm O}$.
Establishing the relationship between $S_{\rm R}$ and $S_{\rm MWO}$ would become straightforward only if we can determine the relationship between $S_{\rm O}$ and $S_{\rm MWO}$.
Since $S_{\rm O}$ is consistent with $S_{\rm R\gtrsim30,000}$ (Section \ref{res.sec}) we can use stars common to high-resolution spectroscopic surveys and the MWO observations to derive these relationship.

\begin{figure}
\includegraphics[width=0.48\textwidth]{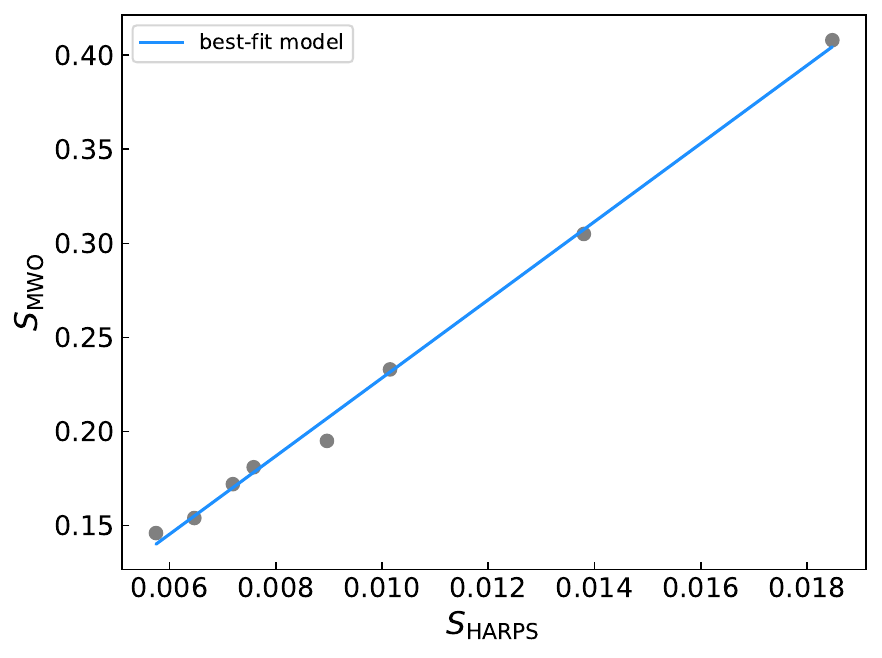}
\caption{Relation between $S_{\rm{HARPS}}$ and $S_{\rm{MWO}}$. Blue line is the best-fit model.}
\label{app}
\end{figure}

Fortunately, there are some common sources between the HARPS spectral observations and the MWO observations \citep{2018A&A...616A.108B}.
The HARPS spectra, with a very high resolution of $R=$ 120,000, allow the $S_{\rm HARPS}$ to be treated as $S_{\rm O}$ without requiring further calibration for resolution effects.
All these spectra were corrected for radial velocities using the values provided by the HARPS data-reduction software \citep{2002Msngr.110....9P}.
For each target, we normalized all the spectra and calculated the $S$-indices of HARPS spectra with $S_{\rm{HARPS}} =(H+K)/(R+V)$.
The median value of these measurements for each source was adopted as the final result (i.e., $S_{\rm{HARPS}}$). We then applied a linear regression to $S_{\rm{HARPS}}$ values and $S_{\rm{MWO}}$ values from \cite{2018A&A...616A.108B} following
\begin{equation}
    S_{\rm{MWO}} = k_{\rm{HARPS}} \times S_{\rm{HARPS}} + b_{\rm{HARPS}}.
\end{equation}
The fitting coefficients are $k_{\rm{HARPS}}=20.767\pm0.515$ and $b_{\rm{HARPS}}=0.021\pm0.005$. Since $S_{\rm{HARPS}}$ were not multiplied by a factor of 8$\alpha$, the fitting results are differ from those reported by \cite{2018A&A...616A.108B}. 
Figure \ref{app} shows the fitting results.

Finally, we provided the general relations between $S_{\rm R}$ and $S_{\rm MWO}$ as follows,
\begin{equation}
\begin{split}
    S_{\rm{MWO}} = k_{\rm{HARPS}} * (k * S_{\rm{R}} + b) + b_{\rm{HARPS}} \\
    = k_{\rm{HARPS}} * k * S_{\rm{R}} + k_{\rm{HARPS}} * b + b_{\rm{HARPS}}\\
    = k^{'} * S_{\rm{R}} + b^{'}.
\end{split}
\end{equation}
The results of $k^{'}$ and $b^{'}$ for each resolution are summarized in table \ref{tab:table1}.

\begin{table*}
\begin{center}
{
\footnotesize
\renewcommand{\arraystretch}{1.}
    \caption{Parameters of the linear fitting}
    \label{tab:table1}
    \begin{tabular}{cccccccc}
    \hline
       Type &
       Resolution & 
       Activity level &
       Stellar Type &
       $k$ &
       $b$ &
       $k^{'}$ &
       $b^{'}$ \\
      \hline
      \multirow{10}{*}{Dwarf (1 \AA)} & \multirow{7}*{1800} & All & All & 2.37 & $-$0.021 & 49.218$\pm$1.221 & $-$0.415$\pm$0.012 \\ \cline{3-8}
                                     & & $S_{\rm R} \geq 0.02$ & All & 2.413 & $-$0.026 & 50.111$\pm$1.243 & $-$0.519$\pm$0.014 \\ \cline{3-8}
                                     & & \multirow{5}*{$S_{\rm R} < 0.02$} & All & 1.461 & $-$0.005 & 30.341$\pm$0.753 & $-$0.083$\pm$0.006 \\  
                                     & & & F & 1.81 & $-$0.014 & 37.588$\pm$0.932 & $-$0.27$\pm$0.009 \\
                                     & & & G & 1.973 & $-$0.015 & 40.973$\pm$1.016 & $-$0.291$\pm$0.009 \\
                                     & & & K & 1.695 & $-$0.007 & 35.2$\pm$0.873 & $-$0.125$\pm$0.006 \\
                                     & & & M & 1.524 & $-$0.006 & 31.649$\pm$0.785 & $-$0.104$\pm$0.006 \\ \cline{2-8}
                                     & \multirow{7}*{2000} & All & All & 2.166 & $-$0.018 & 44.981$\pm$1.116 & $-$0.353$\pm$0.01 \\ \cline{3-8}
                                     & & $S_{\rm R} \geq 0.02$ & All & 2.196 & $-$0.021 & 45.604$\pm$1.131 & $-$0.415$\pm$0.012 \\ \cline{3-8}
                                     & & \multirow{5}*{$S_{\rm R} < 0.02$} & All & 1.435 & $-$0.005 & 29.801$\pm$0.739 & $-$0.083$\pm$0.006 \\  
                                     & & & F & 1.758 & $-$0.013 & 36.508$\pm$0.905 & $-$0.249$\pm$0.008 \\
                                     & & & G & 1.861 & $-$0.013 & 38.647$\pm$0.958 & $-$0.249$\pm$0.008 \\
                                     & & & K & 1.634 & $-$0.006 & 33.933$\pm$0.841 & $-$0.104$\pm$0.006 \\
                                     & & & M & 1.477 & $-$0.005 & 30.673$\pm$0.761 & $-$0.083$\pm$0.006 \\ \cline{2-8}
                                     & 4000 & All & All & 1.381 & $-$0.005 & 28.679$\pm$0.711 & $-$0.083$\pm$0.006 \\ 
                                     & 6000 & All & All & 1.192 & $-$0.002 & 24.754$\pm$0.614 & $-$0.021$\pm$0.005 \\
      \hline
      \multirow{10}{*}{Giant (1 \AA)} & \multirow{7}*{1800} & All & All & 1.641 & $-$0.004 & 34.079$\pm$0.845 & $-$0.062$\pm$0.005 \\ \cline{3-8}
                                     & & $S_{\rm R} \geq 0.01$ & All & 1.734 & $-$0.006 & 36.01$\pm$0.893 & $-$0.104$\pm$0.006 \\ \cline{3-8}
                                     & & \multirow{5}*{$S_{\rm R} < 0.01$} & All & 1.909 & $-$0.005 & 39.644$\pm$0.983 & $-$0.083$\pm$0.006 \\  
                                     & & & F & - & - & - & - \\
                                     & & & G & - & - & - & - \\
                                     & & & K & 2.071 & $-$0.006 & 43.008$\pm$1.067 & $-$0.104$\pm$0.006 \\
                                     & & & M & 1.694 & $-$0.004 & 35.179$\pm$0.872 & $-$0.062$\pm$0.005 \\ \cline{2-8}
                                     & \multirow{7}*{2000} & All & All & 1.646 & $-$0.005 & 34.182$\pm$0.848 & $-$0.083$\pm$0.006 \\ \cline{3-8}
                                     & & $S_{\rm R} \geq 0.01$ & All & 1.755 & $-$0.007 & 36.446$\pm$0.904 & $-$0.125$\pm$0.006 \\ \cline{3-8}
                                     & & \multirow{5}*{$S_{\rm R} < 0.01$} & All & 1.789 & $-$0.004 & 37.152$\pm$0.921 & $-$0.062$\pm$0.005 \\  
                                     & & & F & - & - & - & - \\
                                     & & & G & - & - & - & - \\
                                     & & & K & 1.93 & $-$0.005 & 40.08$\pm$0.994 & $-$0.083$\pm$0.006 \\
                                     & & & M & 1.601 & $-$0.003 & 33.248$\pm$0.825 & $-$0.042$\pm$0.005 \\ \cline{2-8}
                                     & 4000 & All & All & 1.34 & $-$0.003 & 27.828$\pm$0.69 & $-$0.042$\pm$0.005 \\ 
                                     & 6000 & All & All & 1.184 & $-$0.002 & 24.588$\pm$0.61 & $-$0.021$\pm$0.005 \\
      \hline
      \multirow{10}{*}{Giant (2 \AA)} & \multirow{7}*{1800} & All & All & 1.134 & $-$0.002 & 23.55$\pm$0.584 & $-$0.021$\pm$0.005 \\ \cline{3-8}
                                     & & $S_{\rm R} \geq 0.01$ & All & 1.123 & $-$0.002 & 23.321$\pm$0.579 & $-$0.021$\pm$0.005 \\ \cline{3-8}
                                     & & \multirow{5}*{$S_{\rm R} < 0.01$} & All & 1.154 & $-$0.003 & 23.965$\pm$0.594 & $-$0.042$\pm$0.005 \\  
                                     & & & F & - & - & - & - \\
                                     & & & G & - & - & - & - \\
                                     & & & K & 1.264 & $-$0.004 & 26.249$\pm$0.651 & $-$0.062$\pm$0.005 \\
                                     & & & M & 1.02 & $-$0.001 & 21.182$\pm$0.525 & 0.0$\pm$0.005 \\ \cline{2-8}
                                     & \multirow{7}*{2000} & All & All & 1.137 & $-$0.002 & 23.612$\pm$0.586 & $-$0.021$\pm$0.005 \\ \cline{3-8}
                                     & & $S_{\rm R} \geq 0.01$ & All & 1.137 & $-$0.002 & 23.612$\pm$0.585 & $-$0.021$\pm$0.005 \\ \cline{3-8}
                                     & & \multirow{5}*{$S_{\rm R} < 0.01$} & All & 1.12 & $-$0.002 & 23.259$\pm$0.577 & $-$0.021$\pm$0.005 \\  
                                     & & & F & - & - & - & - \\
                                     & & & G & - & - & - & - \\
                                     & & & K & 1.226 & $-$0.003 & 25.46$\pm$0.631 & $-$0.042$\pm$0.005 \\
                                     & & & M & 0.996 & 0.0 & 20.684$\pm$0.513 & 0.021$\pm$0.005 \\ \cline{2-8}
                                     & 4000 & All & All & 1.084 & $-$0.001 & 22.511$\pm$0.558 & 0.0$\pm$0.005 \\ 
                                     & 6000 & All & All & 1.05 & $-$0.001 & 21.805$\pm$0.541 & 0.0$\pm$0.005\\

      \hline
    \end{tabular}\\}
  \end{center}
\end{table*}

\subsection{Application to LAMOST}

\begin{figure}
\includegraphics[width=0.46\textwidth]{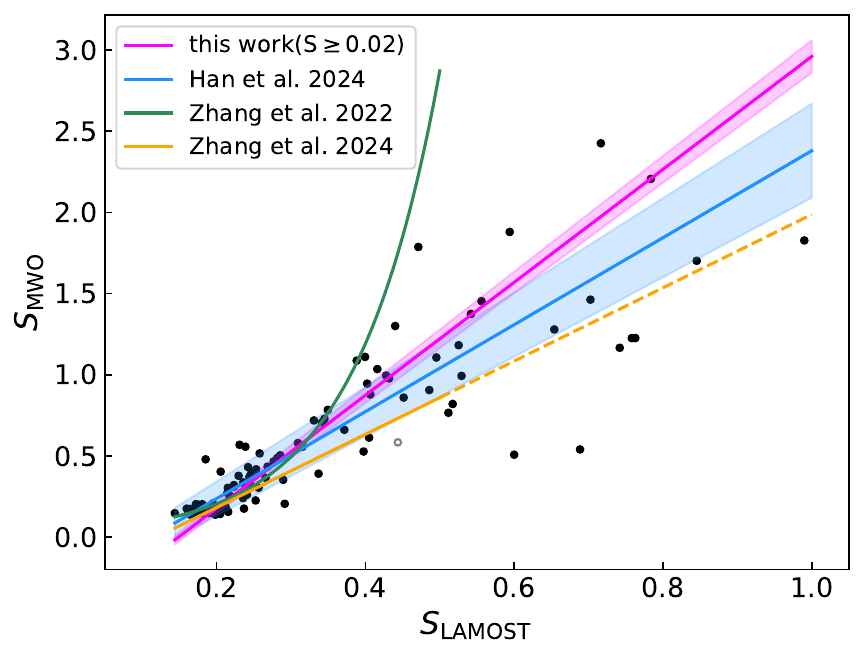}
\caption{Comparison between the different scaling relations converting $S_{\rm{LAMOST}}$ into $S_{\rm{MWO}}$. Open circles represent targets with [Fe/H] values smaller than $-$1. The magenta line is the linear scaling from this work with $k^{'} = 50.111$ and $b^{'} = -0.519$ for dwarfs with $S\geq$0.02 and $\rm R=1800$. Shaded areas are fitting errors. The dotted line represents $S_{\rm MWO} = S_{\rm{LAMOST}}$, meaning no calibration is done as in \citet{2016NatCo...711058K}. Note that here $S_{\rm{LAMOST}} = 8\times1.8\times(H+K)/(R+V)$, while in table \ref{tab:table1}, $S_{\rm R}=(H+K)/(R+V)$.}
\label{comparison}
\end{figure}

The LAMOST low-resolution spectra have been widely used in the studies of stellar magnetic activity \citep[e.g.][]{2020ApJS..247....9Z, 2020ApJ...894L..11Z, 2024ApJ...966...69L, 2024ApJS..271...19Y, 2021ApJS..253...19Z}. 
Some studies \citep[e.g.,][]{2016NatCo...711058K} calculated the $S$-indices as $S_{\rm LAMOST}=8\alpha(H+K)/(R+V)$ and simply treated it as $S_{\rm{MWO}}$ to derive $R_{\rm{HK}}^{'}$ following the method of \cite{2013A&A...549A.117M}.
Although the authors adopted $8\alpha$ as the calibration factor, their approach may introduce inaccuracies in the results.
Furthermore, \cite{2016NatCo...711058K} used the air wavelength to define the line centers of the \cahk lines, whereas the LAMOST spectra are provided in vacuum wavelengths. 
Given the low resolution of LAMOST spectra, this $\approx$1 \AA\ difference could strongly affect the calculation of $S_{\rm{LAMOST}}$.

The most appropriate approach is to first convert $S_{\rm{LAMOST}}$ to $S_{\rm MWO}$ and then calculate $R_{\rm{HK}}^{'}$. 
However, there are no common targets between the LAMOST and MWO observations. Consequently, \citet{2024ApJ...966...69L} and \citet{2024ApJ...977..138H} used common targets between the LAMOST and HARPS observations to do the calibration using the HARPS $S$-indices \citep{2018A&A...616A.108B}, subsequently converting them into $S_{\rm{MWO}}$. 
Similarly, \citet{2022ApJS..263...12Z} used common stars between the LAMOST observations and the $S_{\rm MWO}$ catalog given by \citet{1991ApJS...76..383D}, while \citet{2024ApJS..272...40Z} utilized common targets between the LAMOST observations and several catalogs with $S$-indices calibrated to the MWO measurement.

To compare our results with the conversion relations derived observationally, we plotted the relations from \cite{2022ApJS..263...12Z}, \cite{2024ApJS..272...40Z}, and \cite{2024ApJ...977..138H} in Figure \ref{comparison}.
The coefficients (i.e., slopes and intercepts) of the linear relations are 2.68 and $-0.3$ in \citet{2024ApJ...977..138H} or 2.26 and $-$0.27 for solar-like stars (5400 K $< T_{\rm eff}<$ 6500 K) in \cite{2024ApJS..272...40Z}. 
Note that the LAMOST $S$-indices calculated in these studies were multiplied by a factor of $8\times1.8$, which explains why their coefficients differ from our results in table \ref{tab:table1}.
Additionally, \cite{2022ApJS..263...12Z} used an exponential function to fit the relationship between $S_{\rm{LAMOST}}$ and $S_{\rm{MWO}}$ for solar-like stars. Since both the relations from \cite{2022ApJS..263...12Z} and \cite{2024ApJS..272...40Z} only include solar-like stars with $S_{\rm{LAMOST}} < 0.5$, their relations differ from our results. The magenta line in Figure \ref{comparison} corresponds to the model with $k^{'}=50.111$ and $b^{'}=-0.519$ for dwarfs with $S\geq0.02$ and $\rm R=1800$. The fitting result is different from those from previous literature, especially at large $S_{\rm{LAMOST}}$ region.

The differences of the $S_{\rm{LAMOST}}-S_{\rm{MWO}}$ relations may be caused by the biased sample, since there are only a few targets with large $S$-indices. In previous studies, which provided calibrations from observations, they may be suffered from sample selection bias, low signal-to-noise ratio, and low resolution that could lead to the inaccuracy in the measurement of radial velocities and thus influence the determining of line centers. These effects would lead to large uncertainties while converting $S_{\rm{LAMOST}}$ to $S_{\rm{O}}$ observationally (For example the blue shaded area in Figure \ref{comparison}).

\subsection{Other Effects}

\begin{figure}
\includegraphics[width=0.46\textwidth]{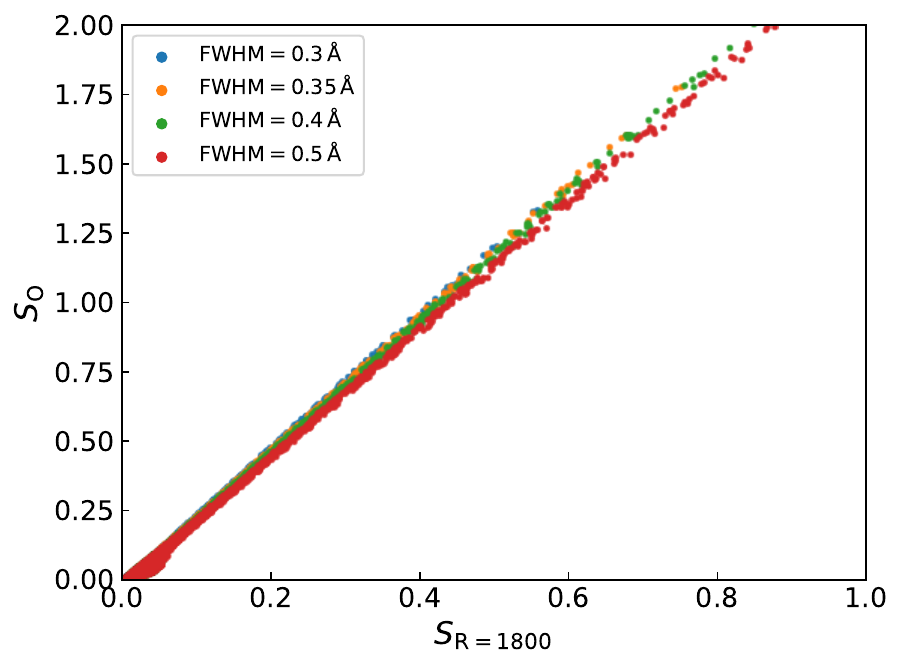}
\caption{Comparison between relations between $S_{\rm{R=1800}}$ and $R_{\rm{O}}$ corresponding to various FWHM of simulated emission cores.}
\label{width}
\end{figure}

First, in this work we used a 0.35 \AA \, Gaussian profile to simulate stellar activity. However, different structures and physical processes in the chromosphere, including the velocity gradients of the fluid, the intensity of shock wave and the non-LTE radiative transfer, would influence the widths of line cores \citep[e.g.][]{1997ApJ...481..500C}. To test the potential influence, in Figure \ref{width} we plot the relations of $S_{\rm{R=1800}}$ and $S_{\rm{O}}$ corresponding to various FWHM of emission cores, i.e., 0.3 \AA, \, 0.35 \AA, \, 0.4 \AA, \, and 0.5 \AA. Obviously, the relations are similar. We conclude that when the simulated emission cores are within the integrated windows, the fitting result will not change significantly. 

Second, the rotational broadening may also have notable impact on the line profile. We further tested the effect using a representative stellar model with $T_{\rm{eff}} = 4900$ K, log$g$ $=$ 4.5 and [Fe/H] $=$ 0.0, corresponding to a K-type dwarfs. Assuming a typical radius of 0.7$R_{\odot}$ and a very short rotation period of 1 day, we derived a FWHM of 0.5 \AA \, of the simulated emission core, which is still smaller than the typical width of integrated window. Thus the rotational broadening will not affect our results significantly.

\section{Summary}

In this study, we examined the influence of spectral resolution on the calculation of $S$-indices using PHOENIX synthetic spectra. We calculated $S_{\rm{O}}$ values at the original resolution of PHOENIX spectra and $S_{\rm{R}}$ values after convolving the spectra to a resolution of 1800, 2000, 3000, 4000, 5000, 6000, 7000, 7500, 8000, 9000, 10,000, 15,000, 20,000, 30,000, 50,000, and 100,000.
For dwarf stars, the $S$-indices were calculated with a 1.09 \AA \, bandpass, while for giant stars, the $S$-indices were calculated in both 1.09 \AA \, bandpass and 2.18 \AA \, bandpass.
Our analysis revealed that lower resolutions led to larger discrepancies between the $S_{\rm{R}}$ and $S_{\rm O}$ values.
Below a spectral resolution of $\approx$30,000, it is necessary to calibrate $S$-indices for accurate estimations. 
For giant stars analyzed within a 2.18 \AA \, window, this threshold shifts to a resolution of $\approx$15,000.

We also provided calibrations from $S_{\rm R}$ to $S_{\rm O}$ values for several current or upcoming spectroscopic surveys, including the LAMOST low-resolution survey with $R\sim1800$, the SEGUE survey with $R\sim2000$, the SDSS-V/BOSS survey with $R\sim2000$, the MUST survey with $R\sim4000$, the DESI survey with $R\sim4000$, and the MSE survey with $R\sim6000$.
These calibrations were conducted specifically for templates with [Fe/H] values higher than $-$1, since templates with lower metallicity do not exhibit a linear relationship between $S_{\rm{R}}$ and $S_{\rm{O}}$.
Additionally, we categorized the templates into high-activity and low-activity groups based on their $S$-indices and developed the scaling relations for each group separately. For low-activity groups, although we divided them into several subgroups, the activity region as well as the differences of their scaling relations are small. Therefore for relations in table \ref{tab:table1}, we recommend to use those with \emph{stellar type = ``All''}.
More importantly, using common targets between the HARPS and MWO observations, we established conversions from $S_{\rm R}$ to $S_{\rm MWO}$, which can be widely applied in studies focusing on chromospheric activity.

\acknowledgements
We thank the referee for the comprehensive suggestions and comments. The Guoshoujing Telescope (the Large Sky Area Multi-Object Fiber Spectroscopic Telescope LAMOST) is a National Major Scientific Project built by the Chinese Academy of Sciences. Funding for the project has been provided by the National Development and Reform Commission. LAMOST is operated and managed by the National Astronomical Observatories, Chinese Academy of Sciences. This work was supported by National Natural Science Foundation of China (NSFC) under grant Nos. 11988101/12273057/11833002/12090042, National Key Research and Development Program of China (NKRDPC) under grant No. 2023YFA1607901, and science research grants from the China Mannned Space Project. J.F.L acknowledges the support from the New Cornerstone Science Foundation through the New Cornerstone Investigator Program and the XPLORER PRIZE.

\bibliographystyle{yahapj}
\bibliography{main}

\begin{thebibliography}{}
\providecommand\natexlab[1]{#1}
\providecommand\JournalTitle[1]{#1}

\bibitem[{{Adibekyan} {et~al.}(2020){Adibekyan}, {Sousa}, {Santos}, {Figueira}, {Allende Prieto}, {Delgado Mena}, {Gonz{\'a}lez Hern{\'a}ndez}, {de Laverny}, {Recio-Blanco}, {Campante}, {Tsantaki}, {Hakobyan}, {Oshagh}, {Faria}, {Bergemann}, {Israelian}, \& {Boulet}}]{2020A&A...642A.182A}
{Adibekyan}, V., {Sousa}, S.~G., {Santos}, N.~C., {et~al.} 2020, \href{http://dx.doi.org/10.1051/0004-6361/202038793}{\JournalTitle{\aap}, 642, A182}

\bibitem[{{Babcock} \& {Babcock}(1955)}]{1955ApJ...121..349B}
{Babcock}, H.~W., \& {Babcock}, H.~D. 1955, \href{http://dx.doi.org/10.1086/145994}{\JournalTitle{\apj}, 121, 349}

\bibitem[{{Bj{\o}rgen} {et~al.}(2018){Bj{\o}rgen}, {Sukhorukov}, {Leenaarts}, {Carlsson}, {de la Cruz Rodr{\'\i}guez}, {Scharmer}, \& {Hansteen}}]{2018A&A...611A..62B}
{Bj{\o}rgen}, J.~P., {Sukhorukov}, A.~V., {Leenaarts}, J., {et~al.} 2018, \href{http://dx.doi.org/10.1051/0004-6361/201731926}{\JournalTitle{\aap}, 611, A62}

\bibitem[{{Boro Saikia} {et~al.}(2018){Boro Saikia}, {Marvin}, {Jeffers}, {Reiners}, {Cameron}, {Marsden}, {Petit}, {Warnecke}, \& {Yadav}}]{2018A&A...616A.108B}
{Boro Saikia}, S., {Marvin}, C.~J., {Jeffers}, S.~V., {et~al.} 2018, \href{http://dx.doi.org/10.1051/0004-6361/201629518}{\JournalTitle{\aap}, 616, A108}

\bibitem[{{Bouchy} {et~al.}(2001){Bouchy}, {Pepe}, \& {Queloz}}]{2001A&A...374..733B}
{Bouchy}, F., {Pepe}, F., \& {Queloz}, D. 2001, \href{http://dx.doi.org/10.1051/0004-6361:20010730}{\JournalTitle{\aap}, 374, 733}

\bibitem[{{Carlsson} \& {Stein}(1997)}]{1997ApJ...481..500C}
{Carlsson}, M., \& {Stein}, R.~F. 1997, \href{http://dx.doi.org/10.1086/304043}{\JournalTitle{\apj}, 481, 500}

\bibitem[{{Cretignier} {et~al.}(2021){Cretignier}, {Dumusque}, {Hara}, \& {Pepe}}]{2021A&A...653A..43C}
{Cretignier}, M., {Dumusque}, X., {Hara}, N.~C., \& {Pepe}, F. 2021, \href{http://dx.doi.org/10.1051/0004-6361/202140986}{\JournalTitle{\aap}, 653, A43}

\bibitem[{{Cretignier} {et~al.}(2024){Cretignier}, {Pietrow}, \& {Aigrain}}]{2024MNRAS.527.2940C}
{Cretignier}, M., {Pietrow}, A.~G.~M., \& {Aigrain}, S. 2024, \href{http://dx.doi.org/10.1093/mnras/stad3292}{\JournalTitle{\mnras}, 527, 2940}

\bibitem[{{Cui} {et~al.}(2012){Cui}, {Zhao}, {Chu}, {Li}, {Li}, {Zhang}, {Su}, {Yao}, {Wang}, {Xing}, {Li}, {Zhu}, {Wang}, {Gu}, {Luo}, {Xu}, {Zhang}, {Liu}, {Zhang}, {Yang}, {Cao}, {Chen}, {Chen}, {Chen}, {Chen}, {Chu}, {Feng}, {Gong}, {Hou}, {Hu}, {Hu}, {Hu}, {Jia}, {Jiang}, {Jiang}, {Jiang}, {Jin}, {Li}, {Li}, {Li}, {Liu}, {Liu}, {Lu}, {Mao}, {Men}, {Qi}, {Qi}, {Shi}, {Tang}, {Tao}, {Wang}, {Wang}, {Wang}, {Wang}, {Wang}, {Wang}, {Wang}, {Wang}, {Wang}, {Wang}, {Wang}, {Wang}, {Xu}, {Xu}, {Yang}, {Yu}, {Yuan}, {Yuan}, {Zhai}, {Zhang}, {Zhang}, {Zhang}, {Zhao}, {Zhou}, {Zhou}, {Zhu}, \& {Zou}}]{2012RAA....12.1197C}
{Cui}, X.-Q., {Zhao}, Y.-H., {Chu}, Y.-Q., {et~al.} 2012, \href{http://dx.doi.org/10.1088/1674-4527/12/9/003}{\JournalTitle{Research in Astronomy and Astrophysics}, 12, 1197}

\bibitem[{{de la Cruz Rodr{\'\i}guez} {et~al.}(2013){de la Cruz Rodr{\'\i}guez}, {De Pontieu}, {Carlsson}, \& {Rouppe van der Voort}}]{2013ApJ...764L..11D}
{de la Cruz Rodr{\'\i}guez}, J., {De Pontieu}, B., {Carlsson}, M., \& {Rouppe van der Voort}, L.~H.~M. 2013, \href{http://dx.doi.org/10.1088/2041-8205/764/1/L11}{\JournalTitle{\apjl}, 764, L11}

\bibitem[{{DESI Collaboration} {et~al.}(2016){DESI Collaboration}, {Aghamousa}, {Aguilar}, {Ahlen}, {Alam}, {Allen}, {Allende Prieto}, {Annis}, {Bailey}, {Balland}, {Ballester}, {Baltay}, {Beaufore}, {Bebek}, {Beers}, {Bell}, {Bernal}, {Besuner}, {Beutler}, {Blake}, {Bleuler}, {Blomqvist}, {Blum}, {Bolton}, {Briceno}, {Brooks}, {Brownstein}, {Buckley-Geer}, {Burden}, {Burtin}, {Busca}, {Cahn}, {Cai}, {Cardiel-Sas}, {Carlberg}, {Carton}, {Casas}, {Castander}, {Cervantes-Cota}, {Claybaugh}, {Close}, {Coker}, {Cole}, {Comparat}, {Cooper}, {Cousinou}, {Crocce}, {Cuby}, {Cunningham}, {Davis}, {Dawson}, {de la Macorra}, {De Vicente}, {Delubac}, {Derwent}, {Dey}, {Dhungana}, {Ding}, {Doel}, {Duan}, {Ealet}, {Edelstein}, {Eftekharzadeh}, {Eisenstein}, {Elliott}, {Escoffier}, {Evatt}, {Fagrelius}, {Fan}, {Fanning}, {Farahi}, {Farihi}, {Favole}, {Feng}, {Fernandez}, {Findlay}, {Finkbeiner}, {Fitzpatrick}, {Flaugher}, {Flender}, {Font-Ribera}, {Forero-Romero}, {Fosalba}, {Frenk}, {Fumagalli}, {Gaensicke}, {Gallo},
  {Garcia-Bellido}, {Gaztanaga}, {Pietro Gentile Fusillo}, {Gerard}, {Gershkovich}, {Giannantonio}, {Gillet}, {Gonzalez-de-Rivera}, {Gonzalez-Perez}, {Gott}, {Graur}, {Gutierrez}, {Guy}, {Habib}, {Heetderks}, {Heetderks}, {Heitmann}, {Hellwing}, {Herrera}, {Ho}, {Holland}, {Honscheid}, {Huff}, {Hutchinson}, {Huterer}, {Hwang}, {Illa Laguna}, {Ishikawa}, {Jacobs}, {Jeffrey}, {Jelinsky}, {Jennings}, {Jiang}, {Jimenez}, {Johnson}, {Joyce}, {Jullo}, {Juneau}, {Kama}, {Karcher}, {Karkar}, {Kehoe}, {Kennamer}, {Kent}, {Kilbingeer}, {Kim}, {Kirkby}, {Kisner}, {Kitanidis}, {Kneib}, {Koposov}, {Kovacs}, {Koyama}, {Kremin}, {Kron}, {Kronig}, {Kueter-Young}, {Lacey}, {Lafever}, {Lahav}, {Lambert}, {Lampton}, {Landriau}, {Lang}, {Lauer}, {Le Goff}, {Le Guillou}, {Le Van Suu}, {Lee}, {Lee}, {Leitner}, {Lesser}, {Levi}, {L'Huillier}, {Li}, {Liang}, {Lin}, {Linder}, {Loebman}, {Luki{\'c}}, {Ma}, {MacCrann}, {Magneville}, {Makarem}, {Manera}, {Manser}, {Marshall}, {Martini}, {Massey}, {Matheson}, {McCauley}, {McDonald},
  {McGreer}, {Meisner}, {Metcalfe}, {Miller}, {Miquel}, {Moustakas}, {Myers}, {Naik}, {Newman}, {Nichol}, {Nicola}, {Nicolati da Costa}, {Nie}, {Niz}, {Norberg}, {Nord}, {Norman}, {Nugent}, {O'Brien}, {Oh}, {Olsen}, {Padilla}, {Padmanabhan}, {Padmanabhan}, {Palanque-Delabrouille}, {Palmese}, {Pappalardo}, {P{\^a}ris}, {Park}, {Patej}, {Peacock}, {Peiris}, {Peng}, {Percival}, {Perruchot}, {Pieri}, {Pogge}, {Pollack}, {Poppett}, {Prada}, {Prakash}, {Probst}, {Rabinowitz}, {Raichoor}, {Ree}, {Refregier}, {Regal}, {Reid}, {Reil}, {Rezaie}, {Rockosi}, {Roe}, {Ronayette}, {Roodman}, {Ross}, {Ross}, {Rossi}, {Rozo}, {Ruhlmann-Kleider}, {Rykoff}, {Sabiu}, {Samushia}, {Sanchez}, {Sanchez}, {Schlegel}, {Schneider}, {Schubnell}, {Secroun}, {Seljak}, {Seo}, {Serrano}, {Shafieloo}, {Shan}, {Sharples}, {Sholl}, {Shourt}, {Silber}, {Silva}, {Sirk}, {Slosar}, {Smith}, {Smoot}, {Som}, {Song}, {Sprayberry}, {Staten}, {Stefanik}, {Tarle}, {Sien Tie}, {Tinker}, {Tojeiro}, {Valdes}, {Valenzuela}, {Valluri}, {Vargas-Magana},
  {Verde}, {Walker}, {Wang}, {Wang}, {Weaver}, {Weaverdyck}, {Wechsler}, {Weinberg}, {White}, {Yang}, {Yeche}, {Zhang}, {Zhao}, {Zheng}, {Zhou}, {Zhou}, {Zhu}, {Zou}, \& {Zu}}]{2016arXiv161100036D}
{DESI Collaboration}, {Aghamousa}, A., {Aguilar}, J., {et~al.} 2016, \href{http://dx.doi.org/10.48550/arXiv.1611.00036}{\JournalTitle{arXiv e-prints}, arXiv:1611.00036}

\bibitem[{{Dineva} {et~al.}(2022){Dineva}, {Pearson}, {Ilyin}, {Verma}, {Diercke}, {Strassmeier}, \& {Denker}}]{2022AN....34323996D}
{Dineva}, E., {Pearson}, J., {Ilyin}, I., {et~al.} 2022, \href{http://dx.doi.org/10.1002/asna.20223996}{\JournalTitle{Astronomische Nachrichten}, 343, e23996}

\bibitem[{{Dumusque} {et~al.}(2021){Dumusque}, {Cretignier}, {Sosnowska}, {Buchschacher}, {Lovis}, {Phillips}, {Pepe}, {Alesina}, {Buchhave}, {Burnier}, {Cecconi}, {Cegla}, {Cloutier}, {Collier Cameron}, {Cosentino}, {Ghedina}, {Gonz{\'a}lez}, {Haywood}, {Latham}, {Lodi}, {L{\'o}pez-Morales}, {Maldonado}, {Malavolta}, {Micela}, {Molinari}, {Mortier}, {P{\'e}rez Ventura}, {Pinamonti}, {Poretti}, {Rice}, {Riverol}, {Riverol}, {San Juan}, {S{\'e}gransan}, {Sozzetti}, {Thompson}, {Udry}, \& {Wilson}}]{2021A&A...648A.103D}
{Dumusque}, X., {Cretignier}, M., {Sosnowska}, D., {et~al.} 2021, \href{http://dx.doi.org/10.1051/0004-6361/202039350}{\JournalTitle{\aap}, 648, A103}

\bibitem[{{Duncan} {et~al.}(1991){Duncan}, {Vaughan}, {Wilson}, {Preston}, {Frazer}, {Lanning}, {Misch}, {Mueller}, {Soyumer}, {Woodard}, {Baliunas}, {Noyes}, {Hartmann}, {Porter}, {Zwaan}, {Middelkoop}, {Rutten}, \& {Mihalas}}]{1991ApJS...76..383D}
{Duncan}, D.~K., {Vaughan}, A.~H., {Wilson}, O.~C., {et~al.} 1991, \href{http://dx.doi.org/10.1086/191572}{\JournalTitle{\apjs}, 76, 383}

\bibitem[{{Gomes da Silva} {et~al.}(2022){Gomes da Silva}, {Bensabat}, {Monteiro}, \& {Santos}}]{2022A&A...668A.174G}
{Gomes da Silva}, J., {Bensabat}, A., {Monteiro}, T., \& {Santos}, N.~C. 2022, \href{http://dx.doi.org/10.1051/0004-6361/202244595}{\JournalTitle{\aap}, 668, A174}

\bibitem[{{Gray} {et~al.}(2006){Gray}, {Corbally}, {Garrison}, {McFadden}, {Bubar}, {McGahee}, {O'Donoghue}, \& {Knox}}]{2006AJ....132..161G}
{Gray}, R.~O., {Corbally}, C.~J., {Garrison}, R.~F., {et~al.} 2006, \href{http://dx.doi.org/10.1086/504637}{\JournalTitle{\aj}, 132, 161}

\bibitem[{{Hall} {et~al.}(2007){Hall}, {Lockwood}, \& {Skiff}}]{2007AJ....133..862H}
{Hall}, J.~C., {Lockwood}, G.~W., \& {Skiff}, B.~A. 2007, \href{http://dx.doi.org/10.1086/510356}{\JournalTitle{\aj}, 133, 862}

\bibitem[{{Han} {et~al.}(2024){Han}, {Wang}, {Li}, {Zheng}, \& {Liu}}]{2024ApJ...977..138H}
{Han}, H., {Wang}, S., {Li}, X., {Zheng}, C., \& {Liu}, J. 2024, \href{http://dx.doi.org/10.3847/1538-4357/ad957a}{\JournalTitle{\apj}, 977, 138}

\bibitem[{{Hill} {et~al.}(2018){Hill}, {Flagey}, {McConnachie}, {Szeto}, {Anthony}, {Ari{\~n}o}, {Babas}, {Bagnoud}, {Baker}, {Barrick}, {Bauman}, {Benedict}, {Berthod}, {Bilbao}, {Bizkarguenaga}, {Blin}, {Bradley}, {Brousseau}, {Brown}, {Brzeski}, {Brzezik}, {Caillier}, {Campo}, {Carton}, {Chu}, {Churilov}, {Crampton}, {Crofoot}, {Dale}, {de Bilbao}, {Sainz de la Maza}, {Devost}, {Edgar}, {Erickson}, {Farrell}, {Fouque}, {Fournier}, {Garrido}, {Gedig}, {Geyskens}, {Gilbert}, {Gillingham}, {Gonz{\'a}lez de Rivera}, {Green}, {Grigel}, {Hall}, {Ho}, {Horville}, {Hu}, {Irusta}, {Isani}, {Jahandar}, {Kaplinghat}, {Kielty}, {Kulkarni}, {Lahidalga}, {Laurent}, {Lawrence}, {Laychak}, {Lee}, {Liu}, {Loewen}, {L{\'o}pez}, {Lorentz}, {Lorgeoux}, {Mahoney}, {Mali}, {Manuel}, {Mart{\'\i}nez}, {Mazoukh}, {Messaddeq}, {Migniau}, {Mignot}, {Monty}, {Morency}, {Mouser}, {Muller}, {Muller}, {Murga}, {Murowinski}, {Nicolov}, {Pai}, {Pawluczyk}, {Pazder}, {P{\'e}contal}, {Petric}, {Prada}, {Rai}, {Ricard}, {Roberts}, {Rodgers},
  {Rodgers}, {Ruan}, {Russelo}, {Salmom}, {S{\'a}nchez}, {Saunders}, {Scott}, {Sheinis}, {Simons}, {Smedley}, {Tang}, {Teran}, {Thibault}, {Thirupathi}, {Tresse}, {Troy}, {Urrutia}, {van Vuuren}, {Venkatesan}, {Venn}, {Vermeulen}, {Villaver}, {Waller}, {Wang}, {Wang}, {Williams}, {Wilson}, {Withington}, {Y{\`e}che}, {Yong}, {Zhai}, {Zhang}, {Zhelem}, \& {Zhou}}]{2018arXiv181008695H}
{Hill}, A., {Flagey}, N., {McConnachie}, A., {et~al.} 2018, \href{http://dx.doi.org/10.48550/arXiv.1810.08695}{\JournalTitle{arXiv e-prints}, arXiv:1810.08695}

\bibitem[{{Husser} {et~al.}(2013){Husser}, {Wende-von Berg}, {Dreizler}, {Homeier}, {Reiners}, {Barman}, \& {Hauschildt}}]{2013A&A...553A...6H}
{Husser}, T.~O., {Wende-von Berg}, S., {Dreizler}, S., {et~al.} 2013, \href{http://dx.doi.org/10.1051/0004-6361/201219058}{\JournalTitle{\aap}, 553, A6}

\bibitem[{{Karoff} {et~al.}(2016){Karoff}, {Knudsen}, {De Cat}, {Bonanno}, {Fogtmann-Schulz}, {Fu}, {Frasca}, {Inceoglu}, {Olsen}, {Zhang}, {Hou}, {Wang}, {Shi}, \& {Zhang}}]{2016NatCo...711058K}
{Karoff}, C., {Knudsen}, M.~F., {De Cat}, P., {et~al.} 2016, \href{http://dx.doi.org/10.1038/ncomms11058}{\JournalTitle{Nature Communications}, 7, 11058}

\bibitem[{{Kollmeier} {et~al.}(2017){Kollmeier}, {Zasowski}, {Rix}, {Johns}, {Anderson}, {Drory}, {Johnson}, {Pogge}, {Bird}, {Blanc}, {Brownstein}, {Crane}, {De Lee}, {Klaene}, {Kreckel}, {MacDonald}, {Merloni}, {Ness}, {O'Brien}, {Sanchez-Gallego}, {Sayres}, {Shen}, {Thakar}, {Tkachenko}, {Aerts}, {Blanton}, {Eisenstein}, {Holtzman}, {Maoz}, {Nandra}, {Rockosi}, {Weinberg}, {Bovy}, {Casey}, {Chaname}, {Clerc}, {Conroy}, {Eracleous}, {G{\"a}nsicke}, {Hekker}, {Horne}, {Kauffmann}, {McQuinn}, {Pellegrini}, {Schinnerer}, {Schlafly}, {Schwope}, {Seibert}, {Teske}, \& {van Saders}}]{2017arXiv171103234K}
{Kollmeier}, J.~A., {Zasowski}, G., {Rix}, H.-W., {et~al.} 2017, \href{http://dx.doi.org/10.48550/arXiv.1711.03234}{\JournalTitle{arXiv e-prints}, arXiv:1711.03234}

\bibitem[{{Leenaarts} {et~al.}(2013){Leenaarts}, {Pereira}, {Carlsson}, {Uitenbroek}, \& {De Pontieu}}]{2013ApJ...772...90L}
{Leenaarts}, J., {Pereira}, T.~M.~D., {Carlsson}, M., {Uitenbroek}, H., \& {De Pontieu}, B. 2013, \href{http://dx.doi.org/10.1088/0004-637X/772/2/90}{\JournalTitle{\apj}, 772, 90}

\bibitem[{{Li} {et~al.}(2024){Li}, {Wang}, {Han}, {Yang}, {Zheng}, {Huang}, \& {Liu}}]{2024ApJ...966...69L}
{Li}, X., {Wang}, S., {Han}, H., {et~al.} 2024, \href{http://dx.doi.org/10.3847/1538-4357/ad3038}{\JournalTitle{\apj}, 966, 69}

\bibitem[{{Luo} {et~al.}(2015){Luo}, {Zhao}, {Zhao}, {Deng}, {Liu}, {Jing}, {Wang}, {Zhang}, {Shi}, {Cui}, {Chu}, {Li}, {Bai}, {Wu}, {Cai}, {Cao}, {Cao}, {Carlin}, {Chen}, {Chen}, {Chen}, {Chen}, {Chen}, {Chen}, {Chen}, {Christlieb}, {Chu}, {Cui}, {Dong}, {Du}, {Fan}, {Feng}, {Fu}, {Gao}, {Gong}, {Gu}, {Guo}, {Han}, {He}, {Hou}, {Hou}, {Hou}, {Hu}, {Hu}, {Hu}, {Huo}, {Jia}, {Jiang}, {Jiang}, {Jiang}, {Jin}, {Kong}, {Kong}, {Lei}, {Li}, {Li}, {Li}, {Li}, {Li}, {Li}, {Li}, {Li}, {Li}, {Li}, {Li}, {Li}, {Liang}, {Lin}, {Liu}, {Liu}, {Liu}, {Liu}, {Lu}, {Luo}, {Mao}, {Newberg}, {Ni}, {Qi}, {Qi}, {Shen}, {Shi}, {Song}, {Song}, {Su}, {Su}, {Tang}, {Tao}, {Tian}, {Wang}, {Wang}, {Wang}, {Wang}, {Wang}, {Wang}, {Wang}, {Wang}, {Wang}, {Wang}, {Wang}, {Wang}, {Wang}, {Wang}, {Wang}, {Wang}, {Wang}, {Wang}, {Wang}, {Wang}, {Wei}, {Wei}, {Wu}, {Wu}, {Wu}, {Wu}, {Xing}, {Xu}, {Xu}, {Xu}, {Yan}, {Yang}, {Yang}, {Yang}, {Yang}, {Yao}, {Yu}, {Yuan}, {Yuan}, {Yuan}, {Yuan}, {Zhai}, {Zhang}, {Zhang}, {Zhang}, {Zhang},
  {Zhang}, {Zhang}, {Zhang}, {Zhang}, {Zhao}, {Zhou}, {Zhou}, {Zhu}, {Zhu}, {Zou}, \& {Zuo}}]{2015RAA....15.1095L}
{Luo}, A.~L., {Zhao}, Y.-H., {Zhao}, G., {et~al.} 2015, \href{http://dx.doi.org/10.1088/1674-4527/15/8/002}{\JournalTitle{Research in Astronomy and Astrophysics}, 15, 1095}

\bibitem[{{Mittag} {et~al.}(2013){Mittag}, {Schmitt}, \& {Schr{\"o}der}}]{2013A&A...549A.117M}
{Mittag}, M., {Schmitt}, J.~H.~M.~M., \& {Schr{\"o}der}, K.~P. 2013, \href{http://dx.doi.org/10.1051/0004-6361/201219868}{\JournalTitle{\aap}, 549, A117}

\bibitem[{{Noyes} {et~al.}(1984){Noyes}, {Hartmann}, {Baliunas}, {Duncan}, \& {Vaughan}}]{1984ApJ...279..763N}
{Noyes}, R.~W., {Hartmann}, L.~W., {Baliunas}, S.~L., {Duncan}, D.~K., \& {Vaughan}, A.~H. 1984, \href{http://dx.doi.org/10.1086/161945}{\JournalTitle{\apj}, 279, 763}

\bibitem[{{Pepe} {et~al.}(2002){Pepe}, {Mayor}, {Rupprecht}, {Avila}, {Ballester}, {Beckers}, {Benz}, {Bertaux}, {Bouchy}, {Buzzoni}, {Cavadore}, {Deiries}, {Dekker}, {Delabre}, {D'Odorico}, {Eckert}, {Fischer}, {Fleury}, {George}, {Gilliotte}, {Gojak}, {Guzman}, {Koch}, {Kohler}, {Kotzlowski}, {Lacroix}, {Le Merrer}, {Lizon}, {Lo Curto}, {Longinotti}, {Megevand}, {Pasquini}, {Petitpas}, {Pichard}, {Queloz}, {Reyes}, {Richaud}, {Sivan}, {Sosnowska}, {Soto}, {Udry}, {Ureta}, {van Kesteren}, {Weber}, {Weilenmann}, {Wicenec}, {Wieland}, {Christensen-Dalsgaard}, {Dravins}, {Hatzes}, {K{\"u}rster}, {Paresce}, \& {Penny}}]{2002Msngr.110....9P}
{Pepe}, F., {Mayor}, M., {Rupprecht}, G., {et~al.} 2002, \JournalTitle{The Messenger}, 110, 9

\bibitem[{{Pietrow} {et~al.}(2024){Pietrow}, {Cretignier}, {Druett}, {Alvarado-G{\'o}mez}, {Hofmeister}, {Verma}, {Kamlah}, {Baratella}, {Amazo-G{\'o}mez}, {Kontogiannis}, {Dineva}, {Warmuth}, {Denker}, {Poppenhaeger}, {Andriienko}, {Dumusque}, \& {L{\"o}fdahl}}]{2024A&A...682A..46P}
{Pietrow}, A.~G.~M., {Cretignier}, M., {Druett}, M.~K., {et~al.} 2024, \href{http://dx.doi.org/10.1051/0004-6361/202347895}{\JournalTitle{\aap}, 682, A46}

\bibitem[{{Reiners}(2009)}]{2009A&A...498..853R}
{Reiners}, A. 2009, \href{http://dx.doi.org/10.1051/0004-6361/200810257}{\JournalTitle{\aap}, 498, 853}

\bibitem[{{Rouppe van der Voort}(2002)}]{2002A&A...389.1020R}
{Rouppe van der Voort}, L.~H.~M. 2002, \href{http://dx.doi.org/10.1051/0004-6361:20020638}{\JournalTitle{\aap}, 389, 1020}

\bibitem[{{Rutten} {et~al.}(2004){Rutten}, {de Wijn}, \& {S{\"u}tterlin}}]{2004A&A...416..333R}
{Rutten}, R.~J., {de Wijn}, A.~G., \& {S{\"u}tterlin}, P. 2004, \href{http://dx.doi.org/10.1051/0004-6361:20035636}{\JournalTitle{\aap}, 416, 333}

\bibitem[{{Schr{\"o}der} {et~al.}(2009){Schr{\"o}der}, {Reiners}, \& {Schmitt}}]{2009A&A...493.1099S}
{Schr{\"o}der}, C., {Reiners}, A., \& {Schmitt}, J.~H.~M.~M. 2009, \href{http://dx.doi.org/10.1051/0004-6361:200810377}{\JournalTitle{\aap}, 493, 1099}

\bibitem[{{Schr{\"o}der} {et~al.}(2018){Schr{\"o}der}, {Schmitt}, {Mittag}, {G{\'o}mez Trejo}, \& {Jack}}]{2018MNRAS.480.2137S}
{Schr{\"o}der}, K.~P., {Schmitt}, J.~H.~M.~M., {Mittag}, M., {G{\'o}mez Trejo}, V., \& {Jack}, D. 2018, \href{http://dx.doi.org/10.1093/mnras/sty1942}{\JournalTitle{\mnras}, 480, 2137}

\bibitem[{{Sheeley}(1967)}]{1967ApJ...147.1106S}
{Sheeley}, Jr., N.~R. 1967, \href{http://dx.doi.org/10.1086/149099}{\JournalTitle{\apj}, 147, 1106}

\bibitem[{{Sheminova}(2012)}]{2012SoPh..280...83S}
{Sheminova}, V.~A. 2012, \href{http://dx.doi.org/10.1007/s11207-012-0066-x}{\JournalTitle{\solphys}, 280, 83}

\bibitem[{{Sowmya} {et~al.}(2023){Sowmya}, {Shapiro}, {Rouppe van der Voort}, {Krivova}, \& {Solanki}}]{2023ApJ...956L..10S}
{Sowmya}, K., {Shapiro}, A.~I., {Rouppe van der Voort}, L.~H.~M., {Krivova}, N.~A., \& {Solanki}, S.~K. 2023, \href{http://dx.doi.org/10.3847/2041-8213/acf92a}{\JournalTitle{\apjl}, 956, L10}

\bibitem[{{Suzuki} {et~al.}(2003){Suzuki}, {Tytler}, {Kirkman}, {O'Meara}, \& {Lubin}}]{2003PASP..115.1050S}
{Suzuki}, N., {Tytler}, D., {Kirkman}, D., {O'Meara}, J.~M., \& {Lubin}, D. 2003, \href{http://dx.doi.org/10.1086/376849}{\JournalTitle{\pasp}, 115, 1050}

\bibitem[{{Vaughan} {et~al.}(1978){Vaughan}, {Preston}, \& {Wilson}}]{1978PASP...90..267V}
{Vaughan}, A.~H., {Preston}, G.~W., \& {Wilson}, O.~C. 1978, \href{http://dx.doi.org/10.1086/130324}{\JournalTitle{\pasp}, 90, 267}

\bibitem[{{Vernazza} {et~al.}(1981){Vernazza}, {Avrett}, \& {Loeser}}]{1981ApJS...45..635V}
{Vernazza}, J.~E., {Avrett}, E.~H., \& {Loeser}, R. 1981, \href{http://dx.doi.org/10.1086/190731}{\JournalTitle{\apjs}, 45, 635}

\bibitem[{{Wilson}(1968)}]{1968ApJ...153..221W}
{Wilson}, O.~C. 1968, \href{http://dx.doi.org/10.1086/149652}{\JournalTitle{\apj}, 153, 221}

\bibitem[{{Wilson}(1978)}]{1978ApJ...226..379W}
{Wilson}, O.~C. 1978, \href{http://dx.doi.org/10.1086/156618}{\JournalTitle{\apj}, 226, 379}

\bibitem[{{Wilson} \& {Vainu Bappu}(1957)}]{1957ApJ...125..661W}
{Wilson}, O.~C., \& {Vainu Bappu}, M.~K. 1957, \href{http://dx.doi.org/10.1086/146339}{\JournalTitle{\apj}, 125, 661}

\bibitem[{{Wright} {et~al.}(2004){Wright}, {Marcy}, {Butler}, \& {Vogt}}]{2004ApJS..152..261W}
{Wright}, J.~T., {Marcy}, G.~W., {Butler}, R.~P., \& {Vogt}, S.~S. 2004, \href{http://dx.doi.org/10.1086/386283}{\JournalTitle{\apjs}, 152, 261}

\bibitem[{{Yanny} {et~al.}(2009){Yanny}, {Rockosi}, {Newberg}, {Knapp}, {Adelman-McCarthy}, {Alcorn}, {Allam}, {Allende Prieto}, {An}, {Anderson}, {Anderson}, {Bailer-Jones}, {Bastian}, {Beers}, {Bell}, {Belokurov}, {Bizyaev}, {Blythe}, {Bochanski}, {Boroski}, {Brinchmann}, {Brinkmann}, {Brewington}, {Carey}, {Cudworth}, {Evans}, {Evans}, {Gates}, {G{\"a}nsicke}, {Gillespie}, {Gilmore}, {Nebot Gomez-Moran}, {Grebel}, {Greenwell}, {Gunn}, {Jordan}, {Jordan}, {Harding}, {Harris}, {Hendry}, {Holder}, {Ivans}, {Ivezi{\v{c}}}, {Jester}, {Johnson}, {Kent}, {Kleinman}, {Kniazev}, {Krzesinski}, {Kron}, {Kuropatkin}, {Lebedeva}, {Lee}, {French Leger}, {L{\'e}pine}, {Levine}, {Lin}, {Long}, {Loomis}, {Lupton}, {Malanushenko}, {Malanushenko}, {Margon}, {Martinez-Delgado}, {McGehee}, {Monet}, {Morrison}, {Munn}, {Neilsen}, {Nitta}, {Norris}, {Oravetz}, {Owen}, {Padmanabhan}, {Pan}, {Peterson}, {Pier}, {Platson}, {Re Fiorentin}, {Richards}, {Rix}, {Schlegel}, {Schneider}, {Schreiber}, {Schwope}, {Sibley}, {Simmons},
  {Snedden}, {Allyn Smith}, {Stark}, {Stauffer}, {Steinmetz}, {Stoughton}, {SubbaRao}, {Szalay}, {Szkody}, {Thakar}, {Sivarani}, {Tucker}, {Uomoto}, {Vanden Berk}, {Vidrih}, {Wadadekar}, {Watters}, {Wilhelm}, {Wyse}, {Yarger}, \& {Zucker}}]{2009AJ....137.4377Y}
{Yanny}, B., {Rockosi}, C., {Newberg}, H.~J., {et~al.} 2009, \href{http://dx.doi.org/10.1088/0004-6256/137/5/4377}{\JournalTitle{\aj}, 137, 4377}

\bibitem[{{Ye} {et~al.}(2024){Ye}, {Bi}, {Zhang}, {Sun}, {Long}, {Ge}, {Li}, {Zhang}, {Chen}, {Li}, {Zhou}, \& {Xiang}}]{2024ApJS..271...19Y}
{Ye}, L., {Bi}, S., {Zhang}, J., {et~al.} 2024, \href{http://dx.doi.org/10.3847/1538-4365/ad1eee}{\JournalTitle{\apjs}, 271, 19}

\bibitem[{{Zhang} {et~al.}(2020{\natexlab{a}}){Zhang}, {Bi}, {Li}, {Jiang}, {Li}, {He}, {Yu}, {Khanna}, {Ge}, {Liu}, {Tian}, {Wu}, \& {Zhang}}]{2020ApJS..247....9Z}
{Zhang}, J., {Bi}, S., {Li}, Y., {et~al.} 2020{\natexlab{a}}, \href{http://dx.doi.org/10.3847/1538-4365/ab6165}{\JournalTitle{\apjs}, 247, 9}

\bibitem[{{Zhang} {et~al.}(2020{\natexlab{b}}){Zhang}, {Shapiro}, {Bi}, {Xiang}, {Reinhold}, {Sowmya}, {Li}, {Li}, {Yu}, {Du}, \& {Zhang}}]{2020ApJ...894L..11Z}
{Zhang}, J., {Shapiro}, A.~I., {Bi}, S., {et~al.} 2020{\natexlab{b}}, \href{http://dx.doi.org/10.3847/2041-8213/ab8795}{\JournalTitle{\apjl}, 894, L11}

\bibitem[{{Zhang} {et~al.}(2024){Zhang}, {Xiang}, {Yu}, {Ge}, {Xie}, {Zhang}, {Li}, {Wu}, {Li}, {Bi}, {Yan}, \& {Shi}}]{2024ApJS..272...40Z}
{Zhang}, J., {Xiang}, M., {Yu}, J., {et~al.} 2024, \href{http://dx.doi.org/10.3847/1538-4365/ad41b6}{\JournalTitle{\apjs}, 272, 40}

\bibitem[{{Zhang} {et~al.}(2021){Zhang}, {Meng}, {Long}, {Shi}, {Zhong}, {Han}, {Misra}, \& {Wang}}]{2021ApJS..253...19Z}
{Zhang}, L.-y., {Meng}, G., {Long}, L., {et~al.} 2021, \href{http://dx.doi.org/10.3847/1538-4365/abd7a8}{\JournalTitle{\apjs}, 253, 19}

\bibitem[{{Zhang} {et~al.}(2022){Zhang}, {Zhang}, {He}, {Song}, {Luo}, \& {Zhang}}]{2022ApJS..263...12Z}
{Zhang}, W., {Zhang}, J., {He}, H., {et~al.} 2022, \href{http://dx.doi.org/10.3847/1538-4365/ac9406}{\JournalTitle{\apjs}, 263, 12}

\bibitem[{{Zhao} {et~al.}(2024){Zhao}, {Huang}, {He}, {Montero-Camacho}, {Liu}, {Renard}, {Tang}, {Verdier}, {Xu}, {Yang}, {Yu}, {Zhang}, {Zhao}, {Zhou}, {He}, {Kneib}, {Li}, {Li}, {Wang}, {Xianyu}, {Zhang}, {Gsponer}, {Li}, {Rocher}, {Zou}, {Tan}, {Huang}, {Wang}, {Li}, {Rombach}, {Dong}, {Forero-Sanchez}, {Shan}, {Wang}, {Li}, {Zhai}, {Wang}, {Zhao}, {Shi}, {Mao}, {Huang}, {Guo}, \& {Cai}}]{2024arXiv241107970Z}
{Zhao}, C., {Huang}, S., {He}, M., {et~al.} 2024, \href{http://dx.doi.org/10.48550/arXiv.2411.07970}{\JournalTitle{arXiv e-prints}, arXiv:2411.07970}

\end{thebibliography}
\clearpage

\end{CJK*}
\end{document}